\def\s2{\sqrt{2}}\def\mn{{\mu\nu}}
\def\Z{{\cal Z}}
\def\X{{X}}\def\calZ{\cal B}
\def\Y{{\cal Y}}\def\H{{\cal H}}
\def\L{{\cal L}}

\def\4est{dx^0\wedge dx^1\wedge dx^2\wedge dx^3 }
\def\interno{\vbox{\hbox{\vbox to .3 truecm{\vfill\hbox to .25 truecm
{\hfill\hfill}\vfill}\vrule}\hrule}\hskip 2pt}

\def\datasotto{\@datatrue\footline={\hfil{\rm \data}\hfil}}
\def\frac#1#2{{#1\over {#2}}}
\def\diff#1#2{{\partial #1\over\partial #2}}
\def\pretitolo{\par\vskip .7cm\noindent}
\def\postitolo{\par\nobreak\vskip .3cm\nobreak}
\def\capo{\par\noindent }

\def\menog{\sqrt {-g}}
\def\det{{\rm det}\ }

\def\due{{1\over 2}}\def\Z{{\cal Z}}

\def\dib#1{\partial _{#1}}
\def\nadib#1{\nabla _{#1}}
\def\nadiba#1{\nabla ^{#1}}

\def\gi#1#2{g_{#1#2}}
\def\gia#1#2{g^{#1 #2}}

\def\ixi#1#2{\xi^{#1 }_{\ #2 }}
\def\id#1#2{\delta ^{#1}_{#2}}
\def\ello{\vbox{\hrule\hbox{\vrule\vbox to 4 pt {\vfill\hbox to
4 pt {\hfill\hfill}\vfill}\vrule}\hrule}}
\def\c{\hfill{\ello}}

\catcode`@=11
\def\b@lank{ }
\newif\if@simboli\newif\if@riferimenti
\newif\if@incima\newif\if@bozze
\def\bozze{\@bozzetrue
\immediate\write16{printing eqnames}}
\newwrite\file@simboli
\def\simboli{\immediate\write16{Writing \jobname.SMB }
\@simbolitrue\immediate\openout\file@simboli=\jobname.smb
\immediate\write\file@simboli{Simboli di \jobname}}
\newwrite\file@ausiliario
\def\riferimentifuturi{
\immediate\write16{\jobname.AUX }
\@riferimentitrue\openin1 \jobname.aux
\ifeof1\relax\else\closein1\relax\input\jobname.aux\fi
\immediate\openout\file@ausiliario=\jobname.aux}
\newcount\eq@num\global\eq@num=0
\newcount\sect@num\global\sect@num=0
\newcount\lemm@num\global\lemm@num=0
\newif\if@ndoppia
\def\numerazionedoppia{\@ndoppiatrue\gdef\la@sezionecorrente{\the\sect@num}}
\def\se@indefinito#1{\expandafter\ifx\csname#1\endcsname\relax}
\def\spo@glia#1>{}
\newif\if@primasezione
\@primasezionetrue
\def\s@ection#1\par{\immediate
\write16{#1}\if@primasezione\global\@primasezionefalse\else\goodbreak
\vskip\spaziosoprasez\fi\noindent
{\bf#1}\nobreak\vskip\spaziosottosez\nobreak\noindent}
\def\sezpreset#1{\global\sect@num=#1
\immediate\write16{ !!! sez-preset = #1 }   }
\def\spaziosoprasez{50pt plus 60pt}
\def\spaziosottosez{15pt}
\def\sref#1{\se@indefinito{@s@#1}\immediate\write16{ ??? \string\sref{#1}
non definita !!!}
\expandafter\xdef\csname @s@#1\endcsname{??}\fi\csname @s@#1\endcsname}
\def\autosez#1#2\par{
\global\advance\sect@num by 1\if@ndoppia\global\eq@num=0\fi
\global\lemm@num=0
\xdef\la@sezionecorrente{\the\sect@num}
\def\usa@getta{1}\se@indefinito{@s@#1}\def\usa@getta{2}\fi
\expandafter\ifx\csname @s@#1\endcsname\la@sezionecorrente\def
\usa@getta{2}\fi
\ifodd\usa@getta\immediate\write16
{ ??? possibili riferimenti errati a \string\sref{#1} !!!}\fi
\expandafter\xdef\csname @s@#1\endcsname{\la@sezionecorrente}
\immediate\write16{\la@sezionecorrente. #2}
\if@simboli
\immediate\write\file@simboli{ }\immediate\write\file@simboli{ }
\immediate\write\file@simboli{  Sezione
\la@sezionecorrente :   sref.   #1}
\immediate\write\file@simboli{ } \fi
\if@riferimenti
\immediate\write\file@ausiliario{\string\expandafter\string\edef
\string\csname\b@lank @s@#1\string\endcsname{\la@sezionecorrente}}\fi
\goodbreak\vskip 48pt plus 60pt
\noindent{\bf\the\sect@num.\quad #2}
\if@bozze{\tt #1}\fi    \par\nobreak\vskip 15pt    \nobreak\noindent}
\def\semiautosez#1#2\par{
\gdef\la@sezionecorrente{#1}\if@ndoppia\global\eq@num=0\fi
\if@simboli
\immediate\write\file@simboli{ }\immediate\write\file@simboli{ }
\immediate\write\file@simboli{  Sezione ** : sref.
\expandafter\spo@glia\meaning\la@sezionecorrente}
\immediate\write\file@simboli{ }\fi
\s@ection#2\par}
\def\eqpreset#1{\global\eq@num=#1
\immediate\write16{ !!! eq-preset = #1 }     }
\def\eqlabel#1{\global\advance\eq@num by 1
\if@ndoppia\xdef\il@numero{(\la@sezionecorrente.\the\eq@num)}
\else\xdef\il@numero{(\the\eq@num)}\fi
\def\usa@getta{1}\se@indefinito{@eq@#1}\def\usa@getta{2}\fi
\expandafter\ifx\csname @eq@#1\endcsname\il@numero\def\usa@getta{2}\fi
\ifodd\usa@getta\immediate\write16
{ ??? possibili riferimenti errati a \string\eqref{#1} !!!}\fi
\expandafter\xdef\csname @eq@#1\endcsname{\il@numero}
\if@ndoppia
\def\usa@getta{\expandafter\spo@glia\meaning
\la@sezionecorrente.\the\eq@num}
\else\def\usa@getta{\the\eq@num}\fi
\if@simboli
\immediate\write\file@simboli{  Equazione
\usa@getta :  eqref.   #1}\fi
\if@riferimenti
\immediate\write\file@ausiliario{\string\expandafter\string\edef
\string\csname\b@lank @eq@#1\string\endcsname{\usa@getta}}\fi}
\def\eqref#1{\se@indefinito{@eq@#1}
\immediate\write16{ ??? \string\eqref{#1} non definita !!!}
\if@riferimenti\relax
\else\eqlabel{#1} ???\fi
\fi\csname @eq@#1\endcsname }
\def\autoeqno#1{\eqlabel{#1}\eqno\csname @eq@#1\endcsname\if@bozze
{\tt #1}\else\relax\fi}
\def\autoleqno#1{\eqlabel{#1}\leqno(\csname @eq@#1\endcsname)}

\def\titoli#1{\@incimatrue\nopagenumbers\xdef\prima@riga{#1}
\if@incima
\voffset=+30pt
\headline={\if\pageno=1{\hfil}\else\hfil{\sl \prima@riga}\hfil\folio\fi}\fi}
\catcode`@=12
\def\m#1{p_{#1}}
\def\r{{\rm r}}
\def\t{{\rm t}}
\def\s{{\rm s}}

\hsize 12cm
\magnification 1200
\nopagenumbers
\capo
\
\vskip .5 truecm
\centerline{{\bf UNCONSTRAINED HAMILTONIAN FORMULATION}}\capo
\centerline{{\bf OF GENERAL RELATIVITY
}}\capo
\centerline{\bf WITH THERMO--ELASTIC SOURCES}

\footnote{\ }{
Class. Quantum Gravity {\bf 15} (1998) p.3891-3916}
\capo
\capo
\vskip .5 truecm

\capo
\centerline{\bf Jerzy Kijowski}\capo
\centerline{Centre for Theoretical Physics, Polish Academy of Sciences}\capo
\centerline{Al. Lotnik\'ow 32/46; 02--668
Warsaw, Poland}
\capo
\vskip .5 truecm
\capo
\centerline{\bf Giulio Magli}\capo
\centerline{Dipartimento di Matematica del Politecnico
di Milano}\capo
\centerline{ Piazza Leonardo da Vinci 32, 20133 Milano, Italy}
\centerline{ Fax 0039-2-23994568 e-mail magli@mate.polimi.it}
\capo
\vskip 2 truecm
\
{\sl
A new formulation of the Hamiltonian dynamics of the gravitational field
interacting with
(non--dissipative) thermo--elastic matter is discussed.
It is based on
a gauge condition which allows us to encode the six degrees of freedom of
the ``gravity + matter''--system (two gravitational and four
thermo-mechanical ones), together with their conjugate momenta, in the
Riemannian metric $q_{ij}$ and its conjugate ADM momentum $P^{ij}$.
These variables are not subject to constraints.  We prove
that the Hamiltonian of this system is equal to the total matter
entropy. It generates uniquely the dynamics once expressed as a
function of the canonical variables.  Any function $U$
obtained in this way must fulfil
a system of three, first order,
partial differential equations of the Hamilton--Jacobi type in the variables
$(q_{ij},P^{ij} )$.  These equations are universal and
do not depend upon the properties of the material: its
equation of state enters only as
a boundary condition.
The well posedness of this problem is proved.
Finally, we prove that for vanishing matter
density, the value of $U$ goes to infinity almost everywhere and
remains bounded only on the vacuum constraints.
Therefore the
constrained, vacuum Hamiltonian (zero on constraints and infinity
elsewhere) can be obtained as the limit of a ``deep potential well''
corresponding to non-vanishing matter.
This unconstrained description of Hamiltonian General Relativity
can be useful in numerical calculations
as well as in the canonical approach to Quantum Gravity.
}
\capo

\vfill\eject

\pretitolo
{\bf 1. Introduction}
\postitolo

General Relativity has been formulated as a Hamiltonian field
theory by Arnowitt, Deser and Misner (1963, to be referred towards as
ADM). The ADM paper contained
not only the vacuum case but
also a more general case of gravity interacting
with Maxwell field and charged point particles.  Since then, the
canonical structure of General Relativity coupled with matter fields
has been widely investigated in the case of perfect fluids and,
more recently, in the case of multi--constituent fluids and superfluids
(Comer \& Langlois 1993, 1994).

It is well known that the gauge invariance of Einstein's theory
implies that Hamiltonian General Relativity is a constrained theory.
There have been many attempts to solve these constraints by imposing
tricky gauge conditions, based e.g. on a certain
``geometric time'' used to parameterize space-time points. In the vacuum
case, however, no satisfactory condition of this type has been found.
In presence of matter it is easier to ``gauge the time variable'' in an
invariant way (e.~g.~by using an extra scalar field, whose value is
dynamically identified with time). Such approaches have an obvious
drawback: not all Cauchy surfaces in spacetime are allowed in
the hamiltonian description but only those which fulfil the gauge
condition.

In the present paper we also use matter to
gauge the time but -- in contrast to the above approaches -- {\it all
the spacelike surfaces} are allowed as Cauchy data because our gauge
condition does not fix the time variable but only its scale: it is
fixed by the thermomechanical state of the matter.

For this purpose we need a formulation of continuum mechanics as a
lagrangian field theory. As far as perfect fluids are concerned,
two different
field--theoretical approaches have ben used:
the
Clebsh--potentials one which leads to the so called
``non--canonical'' Poisson structure (see e.g. Holm 1989), and an
approach which, in contrast, may be called ``canonical''. In the latter
approach,
the dynamics is formulated in terms of three unconstrained
field potentials, which assign to each spacetime point $x$
a point $\xi (x)$ of an abstract, three dimensional ``material
space'' ${\calZ}$, equipped with an appropriate geometric structure
(Kijowski \& Tulczyjew 1979, K\"unzle \& Nester 1984).  All the
physical quantities describing the spacetime configuration of the fluid
may be defined in terms of first derivatives of these three potentials
and the equations of hydrodynamics may be formulated in terms of a
system of 2-nd order, hyperbolic equations imposed on these potentials.
\titoli{Unconstrained Hamiltonian formulation of G.R.
with thermo-elastic sources.}

In a recent paper (Kijowski et al. 1990, to be
referred towards as KSG) it was shown how to generalize the above
approach to thermodynamically sensitive materials by adding a new,
``material time'' variable to the material space.
This variable plays the role of a
potential for the temperature, so that the resulting theory
is described by four potentials.
Such potentials can be re--parameterized in an arbitrary way,
since
a re--parameterization
corresponds merely
to a change of the ``label'' and the ``clock''
attached to each particle of the material.
As a consequence,
the theory of non--dissipative, isentropic fluids can be viewed
as a ``gauge type'', lagrangian field theory.

Coupling fluid sources with gravity and passing to the Hamiltonian
description one obtains a generalization of the ADM formalism
in which the canonical variables are the Riemannian metric on each spacelike
hypersurface and the matter fields, together with their conjugate momenta.
The remarkable fact is that, in this approach, a gauge condition
can be used to obtain an unconstrained
Hamiltonian description in which the matter degrees
of freedom and those pertaining to gravity are both encoded in the metric
and in its conjugate momentum only (KSG).

In the present paper we extend the
above described results to the case of an
arbitrary relativistic continuum (for instance a pre-stressed
elastic solid) in the non-dissipative regime. The Hamiltonian
description of the gravitational field obtained in this way displays
interesting, universal geometric properties
(recently, the same approach has been developed for the Hamiltonian dynamics
of self--gravitating shells (Hajicek \& Kijowski 1998)).

Besides its theoretical interest, an unconstrained
formulation of General Relativity in matter
can be useful in
numerical approaches to the dynamics and the oscillations of strongly collapsed
stars. Indeed, due to a process of crystallization of dense
neutron matter, the crust of neutron stars probably exists in the form
of a solid (see e.g. Shapiro \& Teukolsky 1983, Haensel 1995).
It is, moreover, worth mentioning
that the main problem in the canonical approach
to Quantum Gravity relies in the interpretation
of the constraints in operatorial terms (see e.g. Kuchar 1993).
As we shall see, our approach provides naturally
a limiting procedure which allows us to treat the constrained, vacuum
geometrodynamics as a limiting case of an unconstrained theory with
``infinitesimally light''
matter sources. Therefore, our formulation may prove to be useful also
in this context.

The paper is organized as follows.
In section 2 we give a thoroughly review of a
``gauge-field-theoretical'' formulation
of relativistic continuum mechanics.
This theory deserves, in our opinion, such a review because of its
simplicity, internal beauty and universal properties. These properties
are essential for our purposes.

In section 3 we extend the ``gauge--type''
description to the non--isothermal (non dissipative) case.

In section 4
we give the ADM formulation of the Einstein  field equations
in elastic media, and construct the
corresponding variational principle
in Hamiltonian form.

In section 5
we give the reduction of the theory with respect to the
Gauss--Codazzi constraints.
The main tool used is the
{\it comoving gauge} description of the matter fields.
This gauge consists in choosing the
spacetime coordinates $x^\mu$ equal to the four ``material spacetime''
coordinates $\xi^\alpha$, $\alpha = 0,1,2,3$. The three conditions
$x^a=\xi^a$ $(a=1,2,3)$,
actually mean that we use comoving variables as space coordinates,
whereas the condition $x^0 = \xi^0$ is used to define the time
variable in terms of the temperature of the material.
In this gauge, the six degrees of freedom of the composed
``gravity + matter'' system and their corresponding six
momenta are completely
described by the Riemannian metric $q_{ij}$ of the Cauchy
surface and by its conjugate ADM momentum $P^{ij}$. These data are not
constrained: the Gauss--Codazzi equations play the role of
implicit definition of the lapse function and the shift vector.
We prove that the
Hamiltonian of this system is equal to the total amount of entropy
contained in the material under consideration. However, it must be
expressed in terms of the canonical variables $(q_{ij} ,P^{ij})$. For
this purpose the Gauss--Codazzi equations must be solved with respect
to the lapse and the shift.
In this way the entropy can be
expressed as a function $S = S(X, Y_j , q^{ij})$, where
$X$ and $Y_i$ denote the geometric objects built of $q_{ij}$ and
$P^{ij}$, which ``stand on the left hand side'' of the Gauss--Codazzi
equations (the momentum enters
into the Hamiltonian only {\it via} these objects).
The dynamics generated by $S$
is unique. In particular,
the lapse $N$ and the shift $N^k$ equal the derivatives of $S$
with respect to $X$ and $Y_k$ respectively.

To obtain explicitly
the Hamiltonian one needs to follow the above described algebraic
procedure based on the ``inversion'' of the (former) constraints.
There is, however, an underlying ``differential structure''
which, besides giving an equivalent way to calculate the Hamiltonian,
reveals a rich mathematical content which is {\it universal}
in the sense of being hidden in the Einstein--matter
equations independently from the equation of state.
Indeed, it turns out (section 6) that
not all the functions of ten parameters $S = S(X, Y_j , q^{ij})$
can be obtained
starting from all possible materials, described by
all physically admissible state equations: we prove
that the possible Hamiltonian have to fulfil a system of three first order,
partial differential equations of the Hamilton--Jacobi type.
These equations are universal in that they
do not depend upon the specific matter
taken into consideration, the matter properties being
encoded only in the
boundary value of $S$ corresponding to the seven-dimensional subspace
$\{ Y_i = 0\}$.
Since  vanishing of $Y_i$ implies
vanishing of the shift,
in this particular situation we are in the matter rest--frame
and the seven parameters $(X,q^{ij})$ can be identified with
the energy density and the strain tensor
of the material, so that the function
$S = S(X, 0, q^{ij})$ is the state equation.  For a
given material, the Hamiltonian $S = S(X, Y_j , q^{ij})$ can, therefore,
be obtained by solving the Hamilton--Jacobi system on the
10-dimensional space of parameters $(X, Y_j , q^{ij})$, with the state
equation taken as boundary data.
We prove that this boundary problem is well posed and may be
solved uniquely by the method of characteristics.

The above description of the dynamics of the composed ``gravity +
matter'' system sheds new light also on the constrained vacuum dynamics.
Indeed, consider a family of functions
$S_c(X, Y_j , q^{ij}) := S( X/c , {Y_j}/c, q^{ij} )$
derived from a certain reference function $S = S_1$, describing a
reference material, with $c$ being a positive real number. We prove in
Section 6 that the function $S_c$ fulfils automatically the
Hamilton--Jacobi equations if the reference function $S_1$ does,
and that it describes a material whose energy (mass)
is equal to the rescaled energy of the reference material
(so that the new material is $c$ times lighter - or heavier -
than the reference one).
In the limit $c \rightarrow 0$ the material becomes very
light and its influence on the gravitational field can be neglected.
On the other hand, the Hamiltonian $S_c$ tends to infinity
outside of the vacuum constraint submanifold $\{ X=0, Y_k = 0\}$, due to
convexity properties of the entropy in physically reasonable cases.
Therefore
the vacuum Hamiltonian (zero
on the constraints and infinity elsewhere)
can be viewed as the limit of a sequence
of non-constrained Hamiltonian forming a deep potential well with
the constraint manifold taken as bottom.

Our
Hamilton--Jacobi equations are derived in section 6 using
a somewhat technical theorem.
However, the physical origin of such equations is clear.
Indeed, we prove in section 7
that they are equivalent to the {\it local} conservation of
entropy, i.e. to the vanishing of the heat flow.

Finally, in section 8 we discuss in some detail
the particular case of isotropic elastic sources.
In this case
the function of state does not depend of
the entire strain tensor but only on its three invariants.

\pretitolo
{\bf 2. Relativistic mechanics of continua as a lagrangian field
theory.}
\postitolo

Relativistic hydrodynamics is a well established theory (see e.g. Anile
\& Choquet--Bruhat 1989).
The relativistic description of elastic media is slightly less known. It
has been formulated in many different, equivalent ways. The most
important contributions are probably those due to DeWitt (1962),
Souriau (1964), Hernandez (1970),
Maugin (1971,1977,1978a), Carter \&
Quintana (1972), Glass \& Winicour (1972),
Carter (1973), Cattaneo (1973), Bressan (1978).
For complete bibliography and comparative discussion of the various,
equivalent formulations of the theory we refer the reader to
Maugin (1978b) and Kijowski \& Magli (1997).

In the present paper we use a ``gauge--type'' formulation of
relativistic continuum mechanics, which can be used to describe {\it any}
relativistic material (e.~g.~non-homogeneous, pre-stressed etc.)
in a non-dissipative regime. This
formulation can be considered as an obvious generalization of the
``gauge--type'' theory of relativistic elastic
media (Kijowski \& Magli 1992, 1997)
and is essential for our purposes, since
it is especially well adapted for the Hamiltonian description of
``self-gravitating'' continuum materials.
The non--relativistic counterpart of this approach to continuum mechanics
is known as the Piola (or inverse-motion) description
(see e.g. Truesdell \& Toupin 1960).
For a complete formulation of finite elasticity in a language
close to that of the present paper see Maugin's (1993) book.
\par
In this Section, the pseudo--Riemannian geometry $g_{\mu \nu }$ ($\mu
,\nu =0,1,2,3$, signature $(-,+,+,+)$) of the general--relativistic
spacetime ${\cal M}$ is considered as given {\it a priori} (in
Section 4 it will also become a dynamical quantity). To
formulate the dynamical theory of a continuous material moving in ${\cal
M}$,denote by ${\calZ}$ the collection of all the
idealized points (``molecules'') of the material, organized in an
abstract 3--dimensional manifold, the {\it material space}.  The
spacetime configuration of the material is completely described
by a mapping ${\cal G }:{\cal M} \to {\calZ}$, assigning to each
spacetime point $x$ the material point $\xi$ (the specific ``molecule'')
which passes through this point. Each molecule $\xi \in {\calZ }$
follows, therefore, the spacetime trajectory defined as the inverse
image ${\cal G}^{-1} (\xi ) \subset {\cal M}$. Given a coordinate
system $(\xi ^a)$ ($a=1,2,3$) in ${\calZ}$ and a coordinate system
$(x^\mu )$ in ${\cal M}$, the configuration may, thus, be described
by three fields $\xi ^a =\xi ^a(x^\mu )$ depending on four variables
$x^\mu$. We will show how to formulate the physical laws governing the
mechanical properties of the material in terms of a
system of second order, hyperbolic partial differential equations
imposed on the fields. This way the mechanics of continua becomes
a field theory and we may use its standard tools as variational
principles, Noether theorem, Hamiltonian formulation with the
underlying canonical (symplectic) structure of the phase space of
Cauchy data etc.

As a first step we show that the kinematic quantities
characterizing the spacetime configuration of the material,
like the four-velocity $u^\mu$, the matter current $J^\mu$ and the state of
strain, can be encoded in the first derivatives of the fields.
Consider the tangent mapping ${\cal G}_* :T_x {\cal M} \mapsto T_{\xi
(x) } {\calZ}$, described by the $(3 \times 4)$ -- matrix $(\ixi a\mu
):=\left(\partial _\mu \xi ^a \right)$.
We assume this matrix to have maximal rank
and that its one-dimensional kernel to be
time-like (in fact, the dynamical equations of the theory
prevent the fields from violating these conditions in the future,
once they are fulfilled by the Cauchy data).
Vectors
belonging to the kernel of $(\ixi a\mu )$ are tangent to
the world lines of the material, because the value of $\xi^a$ remains
constant on these lines.
It follows that the velocity field
$u^\mu$ can be defined
as the unique future oriented vector field satisfying the
conditions $u^\mu \ixi a\mu =0$ and $u^\mu u_\mu =-1$.
These four
conditions allow to calculate $u^\mu$ uniquely in terms of the
fields' derivatives and the metric (an explicit formula will be given
below).
\par
Given a spacetime configuration
of the material, consider the push-forward of the
contravariant physical metric $g^{\mu\nu}$ from the spacetime ${\cal M}$ to
the material space ${\calZ}$ (Maugin 1978a):
$$
G^{ab}
:=g^{\mu\nu}\ixi a\mu \ixi b\nu \ .\autoeqno{(A1)}
$$
This tensor is obviously symmetric and positive definite.
It defines, therefore, a (time-dependent) Riemannian metric in
${\calZ}$, carrying the information about the actual distances of
adjacent particles of the material, measured in the local rest frame.
Comparing this metric with an appropriate, pre-existing, geometric
structure of ${\calZ}$, describing the mechanical structure of the
material (like e.~g.~volume rigidity or shape rigidity) we can
``decode'' information about the
local state of strain of the material at each instant of time:
more the structure inherited from spacetime ${\cal M}$ ({\it via} the
tensor $G$) differs from the pre-existing structure of ${\calZ}$, higher
is the state of strain of the material under consideration.

Below we give three different examples of such internal structures of
${\calZ}$, corresponding to fluids, isotropic elastic media and
anisotropic (crystalline) materials,
respectively. These structures {\it are not} dynamical objects of the
theory: they are given {\it a priori} for any specific material. We
stress, however, that the dynamical theory we are going to formulate in
the sequel, is universal and applies to any material, whose
physical properties may be described in terms of an appropriate
geometric structure of ${\calZ}$.
\capo
{\it Examples:}
\capo
{\it 1. Volume structure}
\capo
A 3-form (a scalar density)
$$
\omega = r(\xi^a) \ d\xi ^1 \wedge d\xi ^2 \wedge
d\xi ^3 \ ,\autoeqno{omega}
$$
enables to measure the quantity of matter (number of particles or
{\it moles}) contained in a volume $D \subset {\calZ}$
by integration over $D$.
This ``volume structure''
is sufficient to describe the mechanical properties of a
perfect fluid. Indeed, the ratio between the material's own volume form
$\omega$ and the one inherited from spacetime {\it via} $G^{ab}$,
i.~e.~the number
$$
\rho := r \sqrt{\det G^{ab}} \ ,\autoeqno{numero}
$$
describes the actual density of the material (moles per $cm^3$), measured
in the rest frame. Its inverse $v := 1 / \rho $ is equal to
the local, rest--frame specific volume of the fluid ($cm^3$ per mole).
It contains the complete information about the state
of strain of the fluid.
\capo
{\it 2. Metric structure}
\capo
Elastic materials, displaying not only volume rigidity but also shape
rigidity, are equipped with a Riemannian metric $\gamma _{ab}$, the
{\it material metric}. It describes the ``would be'' rest--frame space
distances between neighbouring ``molecules'', measured in the locally
relaxed state of the material.  (To obtain such a locally relaxed
state, we have to extract an ``infinitesimal'' portion from the bulk of
the material. This way the influence of the rest of the material --
possibly pre-stressed -- is eliminated. Such an influence could
otherwise make the relaxation impossible.)

The state of strain of the material is described by the ``ratio''
between the material metric and the physical metric inherited from
${\cal M}$
{\it via} its actual spacetime configuration.
This
``ratio'' may be measured by the tensor
$$
S_{a}^{\ b} := \gamma_{ac} G^{cb} \ .
$$
The material is locally relaxed at the point $x$ if both structures
coincide at $x$, i.~e.~if the actual, physical distances between
material points in the vicinity of $x$ agree with their material
distances. This happens if and only if the strain tensor
is equal to the identity tensor $\delta_a^b$.

The simplest example of a material metric is obviously the flat,
euclidean metric, corresponding to non--pre--stressed materials. A
material carrying such a metric displays no ``internal'' or ``frozen''
stresses and can be embedded into flat Minkowski space {\it without
generating any strain}. Such an embedding is impossible if the material
metric has a non-vanishing curvature.
Materials corresponding to
curved metrics are,
therefore, {\it pre-stressed}
(in what
follows, no specific assumption about $\gamma_{ab}$ will be
necessary).

Denoting by $u_I$ the amount of internal energy (per mole of the
material) of the elastic deformations, accumulated in an infinitesimal
portion of the material during the deformation from the locally relaxed
state to the actual state of strain. It is obvious that, for isotropic
media, this function may depend on the deformation only {\it via} the
invariants of the strain tensor.
Since
the metric $\gamma$ carries automatically a volume structure $r:=
\sqrt{\det \gamma}$, we can take as one of these invariants the
rest frame matter density $\rho$ (or its
inverse $v$), defined exactly as for fluids:
$$
\rho = \sqrt{\det \gamma} \  \sqrt{\det G^{ab}}
= \sqrt{\det S_{a}^{\ b} } \ .\autoeqno{rhov}
$$
As the remaining two invariants of $S$ we can take e.~g.~its trace and
the trace of its square:
$$
h =  S_{a}^{\ a} \ ,\
q =  S_{a}^{\ b} S_{b}^{\ a}  .\autoeqno{inv}
$$
The physical meaning of these invariants
is easily recognized if one considers a weak--strain limit
(Hookean approximation).
In this case the function $u_I$ coincides with the standard formula of
linear elasticity
$$
u_I=\lambda (v) h^2 +2\mu (v) q
$$
where $\lambda$ and $\mu$ are the Lam\'e coefficients,
and $h$ and $q$ are the linear and the quadratic invariants of the
strain, respectively.

\capo
{\it 3. Privileged deformation axis}
\capo
For an anisotropic material (like a crystal) the energy of a
deformation may depend upon its orientation with respect to a specific
axis, reflecting the microscopic composition of the material. The
information about the existence of such an axis
may be encoded in a vector field $E^a$
``frozen'' in ${\calZ}$. We may, therefore, admit an additional
dependence of the energy $u_I$ upon the orientation of $G$ with respect
to one or several vectors $E^a$, i.~e.~upon the quantities
$(G^{-1})_{ab} E^a E^b $.
\par
\vskip .2 truecm
To give an explicit formula for the velocity $u^\mu$ in terms of the
fields $\xi^a$, consider the
pull--back of the material volume form from the material space to
the spacetime. This pull-back is a differential 3--form in the
4--dimensional manifold ${\cal M}$, i.~e.~a vector density $J$ which we
call the {\it material current}. We have
$$
J:={\cal G}^*\omega =
r(\xi ) \ d\xi ^1 (x)\wedge d\xi ^2 (x)\wedge
d\xi ^3 (x)=
r(\xi ) \ \ixi 1\nu \ixi 2\mu \ixi 3\sigma
\ dx^\nu\wedge dx^\mu\wedge dx^\sigma
\ ,
$$
On the other hand, every 3--form in ${\cal M}$ may be written in
the ``vector--density'' representation as
$$
J =J^\mu (\partial_\mu\interno \4est ) \ . \autoeqno{J-mu}
$$
This gives us the following formula for the components $J^\mu$ in
terms of the fields $\xi$ and their derivatives:
$$
J^\mu = r(\xi )\ \epsilon^{\mu\nu\rho\sigma}
\ixi 1\nu \ixi 2\rho \ixi 3\sigma \ ,\autoeqno{sigma}
$$
(here we denote by $\epsilon^{\mu\nu\rho\sigma}$
the standard Levi-Civita tensor density).
The vector density $J$ is {\it a priori} conserved due to its geometric
construction.
Indeed, the exterior derivative of $J$ is equal to the pull--back of
the exterior derivative of $\omega $, and the latter vanishes identically
being a 4--form in the 3--dimensional space $\calZ$:
$$
(\partial _\mu J^\mu ) \4est =
dJ=d({\cal G}^*\omega)={\cal G}^* (d\omega )=0\ ,\autoeqno{dij}
$$
or, equivalently:
$$
\partial _\mu J^\mu =0 \ .
$$
\par
Observe now that
$J^\mu \ixi a\mu  \equiv 0 \ ,$
since $\epsilon^{\mu\nu\rho\sigma}
\ixi 1\nu \ixi 2\rho \ixi 3\sigma \ixi a\mu $ is
the determinant of a matrix with two identical columns.
This means that $J^\mu $ is proportional to the velocity field
and, therefore, may be written in the standard form:
$$
J^\mu =\menog \ \rho u^\mu \ ,\autoeqno{Jdef}
$$
where the scalar $\rho =\sqrt {J^\mu J_\mu /g}$,
with $J^\mu $ given by \eqref{sigma},
is a nonlinear function of $\ixi a\mu$.
Dividing $J^\mu$ given in
\eqref{sigma} by ``$\menog \ \rho$'' defined above, we obtain an explicit
formula for $u^\mu$ in terms of
$\ixi a\mu$.

Being a scalar, the quantity $\rho$ can be calculated
i.~g.~in the material rest frame, where
$u^\mu=(1/\sqrt{-g_{00}},0,0,0)$. In this
frame we have $\ixi a0 = 0$ and, therefore,
$$
\rho = {{J^0} \over {u^0 \menog}} = { { r \  \det
( \ixi ak ) \sqrt{- \det \gia \mu\nu} } \over {\sqrt{-g_{00}}} } =
{ r \  \det
( \ixi ak ) \sqrt{ \det \gia kl} } =
r \ \sqrt{\det G^{ab} }\ .
$$
This proves that the quantity
$\rho$ defined this way coincides, indeed, with the rest
frame matter density defined by \eqref{numero}.

\par
In the present approach, the dynamical equations governing the
evolution of the material under consideration can be derived from the
lagrangian density $\Lambda := -\sqrt{-g}\ \epsilon = -\sqrt{-g}\ \rho
e$, where $\epsilon=\rho e$ denotes the rest frame energy per unit
volume of the material and $e$ denotes the molar rest frame
energy. The mechanical properties of each specific material are completely
encoded in the
function $e = e (G^{ab})$, which describes the dependence of its
energy upon its state of strain. This function plays, therefore, the role
of {\it
equation of state} of the material.
According to the general principles of relativity theory, it must contain
also the molar rest mass $m$, i.e. we have $e =
m+u_I$. In generic situations,
the equation of state
depends also upon the
point $\xi$ {\it via} volume structure, metric structure,
specific deformation axis or any other structure,
which one may find necessary to describe the specific
physical properties of the material. By abuse of
notation we will, however, write $e = e (G^{ab})$
(instead of $e = e (\xi^a , G^{ab})$) whenever it
does not lead to any misunderstanding.
The density $\Lambda$ is, therefore, a first order Lagrangian,
depending upon the unknown fields $\xi ^a$, their first derivatives $\ixi
a\mu $ (which enter through $G^{ab}$)
and -- possibly -- the independent variables $x^\mu $ (which enter {\it
via} the components $g_{\mu\nu}$ of the spacetime metric).  The
dynamical equations of the theory are, thus second order
Euler--Lagrange equations and may be written as follows:
$$
\dib \mu p^\mu_{\ a}=\diff {\Lambda}{\xi^a}\ ,\autoeqno{elag}
$$
where we have introduced the momentum canonically conjugate to
$\xi^a$:
$$
p^\mu_{\ a}:=\diff{\Lambda}{\ixi a\mu} \ , \autoeqno{Piola}
$$
(for historical reasons we may call it the Piola--Kirchhoff momentum
density).

The following identities may be immediately checked in the framework of
the above theory (cfr. Kijowski \& Magli 1992, 1997):
\capo
{{\bf Proposition 1. (Belinfante -- Rosenfeld identity)}}
\capo
{\it The canonical energy-momentum tensor-density}
$$
- {\cal T}^\mu_\nu := p^\mu_{\ a}
\ixi a\nu - \id \mu\nu \Lambda   \ ,\autoeqno{candef}
$$
{\it coincides with the symmetric energy-momentum tensor-density, i.e.
the
following identity holds:}
$$
{\cal T}_{\mu\nu} \equiv  - 2 \diff{\Lambda}{g^{\mu\nu}} \ .
$$
(This identity is a straightforward consequence of the relativistic
invariance of $\Lambda$. It may be checked explicitly by inspection if
we take into account that both $\ixi a\mu$ and
$g_{\mu\nu}$ enter into $\Lambda$ through their combination
\eqref{(A1)} only).
\c

\par
\capo
{{\bf Proposition 2. (Noether identity)}}
\capo
$$
- \nabla _{\mu } {\cal T}^{\mu }_{\nu } \equiv \left(
\partial_\mu\diff \Lambda{\ixi a\mu } -\diff {\Lambda }{\xi^a}\right)
\ixi a\nu
\ .\autoeqno{(6.var4)}
$$
\capo
{\it Proof}:
\capo
Differentiating \eqref{candef} we obtain
$$
- \partial_\mu {\cal T}^{\mu }_{\nu } =
\left( \partial_\mu p^\mu_{\ a} \right) \ixi a\nu +
p^\mu_{\ a}  \ixi a{\nu\mu} - \partial_\nu \Lambda \ .
$$
But
$$
\partial_\nu \Lambda = \diff {\Lambda }{\xi^a}  \ixi a\nu +
\diff \Lambda{\ixi a\mu } \ixi a{\mu\nu} +
\diff{\Lambda}{g^{\sigma\kappa}} \diff{g^{\sigma\kappa}}{x^\nu} \ .
$$
Taking into account the definition \eqref{Piola} of momenta, the
symmetry of the second derivatives ($\ixi a{\nu\mu} \equiv \ixi a{\mu\nu}$)
and the Belinfante -- Rosenfeld identity, we obtain:
$$
- \partial_\mu {\cal T}^{\mu }_{\nu } =
\left(
\partial_\mu\diff \Lambda{\ixi a\mu } -\diff {\Lambda }{\xi^a}\right)
\ixi a\nu + \due {\cal T}_{\sigma\kappa} \diff{g^{\sigma\kappa}}{x^\nu} \ .
$$
Expressing the derivatives of the metric in terms of the connection
coefficients we see that the last term gives exactly the contribution
which is necessary to convert the partial derivative on the left hand
side into the covariant derivative. This ends the proof.\c

We stress that the above identities are purely kinematical. They hold
also for configurations which {\it do not} fulfil the dynamical equations.
In particular, the Noether identity proves that the latter are actually
{\it equivalent} to the energy-momentum conservation $\nabla _{\mu }
{\cal T}^{\mu }_{\nu } = 0$. Indeed, the right hand side of
\eqref{(6.var4)} is automatically orthogonal to the matter velocity
$u^\nu$.
This observation reduces the number of {\it independent} conservation
laws from four down to three -- exactly the number of the
Euler-Lagrange equations.

To see, therefore, that the above theory describes correctly the laws
of continuum mechanics, it is sufficient to calculate the energy-momentum
tensor density ${\cal T}$ and to identify it with the
energy-momentum carried by the material under consideration. For this
purpose we define, at each point of ${\calZ}$ separately, the
{\it response tensor} of the material
$$
Z_{ab} := 2\diff {e}{G^{ab}} \ , \autoeqno{Z-G}
$$
or, equivalently
$$
de (G^{ab} ) = \due \ Z_{ab} \ dG^{ab} \ . \autoeqno{energia-G}
$$

As an example consider an isotropic elastic material, whose energy depends
only upon the
invariants $(v,h,q)$ of the strain. Consequently, the response
tensor may be fully characterized by the following response parameters
$$
p=-\diff ev\ , \
B=\frac 2v \diff eh\ ,\
C=\frac 2v \diff eq\ ,
\autoeqno{response}
$$
according to the formula
$$
Z_{ab}
=v\left(p\ (G^{-1})_{ab} +B \gamma_{ab} + C G_{ab}
\right) \ .
$$
The generating formula \eqref{energia-G} reduces, in this case, to
$$
de(v,h,q) =-p dv + \due v B dh + \frac 12 v C dq  \ .\autoeqno{energia}
$$
The response parameters defined above
describe the reaction of the material to the strain.
In particular, $p$ describes the isotropic stress
while $B$ and $C$ describe the anisotropic response as in the ordinary,
non relativistic elasticity.
The particular case of perfect fluid materials,
corresponding to a constitutive function $e$
which depends only on the specific volume $v$, may be characterized by
the vanishing of both the
anisotropic responses, i.e. by equations $B=C \equiv
0$. Consequently, the response tensor for fluids is proportional to the
physical metric $(G^{-1})_{ab}$ and the generating formula
\eqref{energia-G} reduces to the Pascal law:
$de (v) = - p dv$.

For a general (not necessarily isotropic) material we have the
following:
\capo
{{\bf Proposition 3}}
\capo
{\it
The energy-momentum tensor-density of the above field theory is equal to:}
$$
{\cal T}_\mn =\menog \ \rho \left (e \ u_\mu u_\nu
+  z_{\mu\nu} \right)
\ ,\autoeqno{tenso}
$$
{\it where $z_{\mu\nu}$
is the pull-back of the response tensor $Z_{ab}$
from ${\calZ}$ to ${\cal M}$:}
$$
z_{\mu\nu} := Z_{ab} \ixi a\mu \ixi b\nu \ . \autoeqno{stress}
$$
\capo
{\it Proof:}
\capo
We have:
$$
{\cal T}_{\mu\nu} =2 \diff{\ }{g^{\mu\nu}}
\ \left(\menog \rho e\right)\ .\autoeqno{j}
$$
But:
$$
\eqalign{
\diff{\menog}{\gia \mu\nu}
&= -\due \menog \gi\mu\nu \ ,\cr
\diff{\rho}{\gia \mu\nu}
&=\frac 1{2\rho}
\diff{\ }{\gia \mu\nu}\left(\frac {J^\rho J^\sigma \gi \rho\sigma}{g}
\right)
=\due \rho (\gi \mu\nu +u_\mu u_\nu )\ ,\cr
\diff{e}{\gia \mu\nu}
&=\diff e{G^{ab}}\diff {G^{ab}}{\gia \mu\nu}
=\due Z_{ab} \diff{\left(
\gia \rho\sigma \ixi a\rho \ixi b\sigma\right)}{\gia \mu\nu}
=\due z_{\mu\nu}\ .\cr
}
\autoeqno{wert}
$$
Inserting the above results in \eqref{j} we obtain
eq. \eqref{tenso}.\c
\par
We recognize in formula \eqref{tenso}
the standard energy momentum-tensor of continuum
mechanics, composed of two components: the energy component
``$\epsilon \ u_\mu u_\nu$'', proportional to the velocity, and the
{\it stress tensor} $\rho z_{\mu\nu}$ which is automatically orthogonal to
the velocity. Equations \eqref{stress} (or \eqref{Z-G})
give the stress in terms of the strain ({\it stress --
strain relations}) are uniquely implied by the constitutive equation $e =
e(G^{ab})$ of the material.
\par

The above formulation of continuum mechanics is, of course, invariant with
respect to re\-pa\-rameter\-izations of the material space. Such
reparameterizations may be interpreted as gauge transformations of the
theory. They form the group of all the diffeomorphisms of ${\calZ}$.
Physically, such transformations
consist in changing merely the ``labels'' $\xi^a$
assigned to the molecules of the material. Correspondingly, the fields
$\xi^a$ may be regarded as gauge potentials for the ``elastic field
strength'' $G^{ab}$ which is already gauge invariant. The three
gauge potentials describe the three degrees of freedom of the material.

In Section 4 we are going to include also the gravitational
field as a dynamical quantity.  The group of gauge transformations of the
entire theory (elasticity interacting with gravity) will be the product
of the group of space-time diffeomorphisms (which is the gauge group of
general relativity) by the group of diffeomorphisms of the material
space.
\par

\pretitolo
{\bf 3. Thermodynamics of isentropic flows.}
\postitolo

It is relatively easy to extend the above theory to the thermodynamics
of isentropic flows (no heat conductivity!). For this purpose we
begin with the generating formula
$$
de (G^{ab} , S ) = \due \ Z_{ab} \ dG^{ab} + T dS \ ,\autoeqno{e-term}
$$
which generalizes \eqref{Z-G} to the case of ``thermodynamically
sensitive'' materials. Performing the Legendre transformation
$TdS=d(TS)-SdT$, we obtain an equivalent formula with the Helmholtz free
energy $f:=e-TS $ playing the role of the generating function:
$$
df(G^{ab} ,T) = \due \ Z_{ab} \ dG^{ab} -SdT\ . \autoeqno{f-term}
$$
In the particular case of perfect fluids the above formulae
reduce to $de (v, S ) = -pdv + T dS$ and to
$df(v,T) = -pdv-SdT$, respectively.
\par
Equation \eqref{f-term} suggest that the
the temperature $T$ may be interpreted as a strain and the entropy
as the corresponding stress.
This goes far beyond a formal analogy since it is possible to
express the temperature in terms of derivatives
of a new potential ($\xi^0$, say),
corresponding to a new, time-like dimension - the ``material time'' - in
the material space.
The configurations of the material turn out,
therefore, to be described by {\it four} fields $\xi^\alpha =
\xi^\alpha (x^\mu)$ ($\alpha =0,1,2,3$),
and the ``thermal strain $T$'' can be described in a way
similar to that given by \eqref{(A1)} for the elastic strain.
The required formula for the temperature is the following
(Kijowski \& Tulczyjew 1982, KSG):
$$
T=\beta u^\lambda \ixi 0\lambda \ ,\autoeqno{deftem}
$$
where $\beta$ is a dimension--fixing constant.
\par
Ansatz \eqref{deftem} can be viewed at a purely phenomenological level.
Indeed, in the case of fluids the field
theory derived from the Lagrangian $\Lambda
=-\sqrt{-g}\ \rho f$ (where $f=f(v,T)$ stands for the molar free energy)
describes correctly the relativistic hydrodynamics of
isentropic flows (Kijowski \& Tulczyjew 1982),
and we shall show below that the same
ansatz works for a generic elastic material as well.
However, the potential $\xi^0$ has also a natural
microscopic interpretation
as the retardation of the proper
time of the molecules
with respect to the physical time calculated
over averaged spacetime trajectories of
the idealized continuum material.
Indeed, consider the mean kinetic energy
of the motion of the molecules
of the material, calculated with respect to its rest frame.
For temperatures not too high and average velocities ${\bf v}$
much smaller than one (i.~e.~than the velocity of light) this energy
equals $(1/2)m{\bf v}^2=(3/2)\kappa T$ where $\kappa$ is the Boltzmann
constant.
Consequently,
the proper time $\tau$ of the particles is retarded with respect
to the ``physical'' time $x^0$
(the affine parameter along the tangent to $u^\mu$)
according to formula:
$$
\tau
=\int\sqrt{1-{\bf v}^2}dx^0\approx
\left(1-\frac 32 \frac {\kappa T}m\right)x^0\ .
$$
It follows that,
defining the retardation $\xi^0=x^0-\tau$, we have
$$
T=\beta \diff {\xi^0}{x^0}\ ,
$$
where $\beta=2m/3\kappa$.
Passing from the material rest frame to a general frame
we get \eqref{deftem}.
\par
To show that the Lagrangian
$$
\Lambda = -\menog \rho f(G^{ab} ,T)\ .
$$
describes correctly the thermo--mechanical behaviour of the material,
observe that
this is again a first order Lagrangian which depends upon the first
derivatives $\ixi \alpha\mu$ of the potentials {\it via} the
mechanical strains and the temperature.
Now, we have four independent
Euler-Lagrange equations corresponding to the variation with respect to
four potentials:
$$
\dib \mu P^\mu_{\ \alpha}=\diff {\Lambda}{\xi^\alpha}\ ,\autoeqno{elag-4}
$$
where the momenta canonically conjugate to $\xi^\alpha$ (generalized
Piola--Kirchhoff momenta) are defined as usual:
$$
P^\mu_{\ \alpha}:=\diff {\Lambda}{\ixi \alpha\mu} \ .
$$
Again, we have the following
\capo
{{\bf Proposition 4}}
\capo
{\it Both the Belinfante -- Rosenfeld identity:}
$$
 {\cal T}^\mu_\nu :=  - \left( P^\mu_{\ \alpha}
\ixi \alpha\nu - \id \mu\nu \Lambda \right)   \equiv
- g^{\mu\sigma}  \left( 2 \diff{\Lambda}{g^{\sigma\nu}} \right)
\ ,\autoeqno{BR}
$$
{\it and the Noether identity:}
$$
- \nabla _{\mu } {\cal T}^{\mu }_{\nu } \equiv \left(
\partial_\mu\diff \Lambda{\ixi \alpha\mu } -
\diff {\Lambda }{\xi^\alpha}\right)
\ixi \alpha\nu
\ ,\autoeqno{Noe}
$$
{\it are valid for any field configuration of the above theory, not
necessarily fulfilling the field equations.}
(The proof is an obvious generalization of the previous proofs.)
\par
The Noether identity implies equivalence between the dynamical
equations \eqref{elag-4} of the theory and the
energy-momentum conservation because the deformation gradient $\ixi
\alpha\nu$ is a ($4 \times 4$)-non-degenerate matrix.
It is, however, worthwhile
to notice that the variation with respect to $\xi^0$ produces
simply the entropy conservation law. Indeed, due to \eqref{f-term} and
\eqref{deftem} we have:
$$
P^\mu_{\ 0} =
-\menog \rho \diff {f}T
\diff T{\ixi 0\mu} =
\menog \beta S \rho u^\mu = \beta  S J^\mu  \ .\autoeqno{star}
$$
Because $\xi^0$ does not enter into the Lagrangian, the corresponding
Euler-Lagrange equation reads
$$
0 = \partial_\mu P^\mu_{\ 0} = \partial_\mu (\beta S J^\mu ) =
\beta J^\mu \partial_\mu S \ ,
$$
which means that the amount of entropy contained in each portion of the
material remains constant during the evolution.

The physical interpretation of the remaining three Euler-Lagrange
equations is given by the following:
\capo
{{\bf Proposition 5}}
\capo
{\it The energy-momentum tensor of the above theory is given by the same
formula \eqref{tenso}, where the function
$e$ is defined by the Legendre transformation from \eqref{f-term} back
to \eqref{e-term}, i.e. by the formula $e:= f + TS$.}
\capo
{\it Proof:}
\capo
Due to the Rosenfeld--Belinfante identity we have
$$
{\cal T}^\mu_\nu =2 \diff{\ }{\gia \mu\nu}
\ \left (\menog \rho f\right)\ .
$$
Calculating the above derivative we obtain the same terms
as in the proof of proposition 3 and, moreover, a term arising from the
dependence of the Lagrangian upon $T$. Therefore, we have:
$$
{\cal T}^\mu_\nu =
\menog \rho (fu_\mu u_\nu + z_{\mu\nu}) +2 \menog \rho \diff fT
\diff T{\gia \mu\nu}\ .\autoeqno{numero1}
$$
Now
$$
\diff T{\gia \mu\nu}=
\diff {\ }{\gia \mu\nu}
\left(\frac {\beta J^\lambda \ixi 0\lambda}{\menog \rho}\right)
=\beta J^\lambda \ixi 0\lambda
\diff {\ }{\gia \mu\nu}\frac 1{\menog\rho}
=-\due T u_\mu u_\nu\ ,
$$
where the first two formulae of \eqref{wert} have been used.
Inserting the above result in \eqref{numero1}
and recalling that $\partial f/\partial T = - S$ we obtain formula
\eqref{tenso} with $f+ST$ playing the role of $e$.
\c
\par
The above formulation of relativistic mechanics of continua
as a lagrangian field theory leads in a natural way to
its Hamiltonian counterpart.
In the Hamiltonian formalism
the infinite-dimensional phase space
of Cauchy data for the fields on a given Cauchy surface $\{ t =
{\rm const} \}$ is described by the four configurations variables
$\xi^\alpha$ and their canonical conjugate
momenta $\pi_\alpha$, defined as
derivatives of the Lagrangian with respect to the ``velocities'' ${\dot
\xi^\alpha} = \ixi \alpha{0}$:
$$
\pi_\alpha:=\diff \Lambda{\ixi \alpha{0}}=
P^0_{\ \alpha} \ .\autoeqno{pialfa}
$$
The Poisson bracket between configurations and momenta
assumes its canonical, delta--like form:
$$
\{ \pi_\beta (x) , \xi^\alpha (y) \} = \delta^\alpha_\beta \delta (x,y) \ .
$$
Assuming eqs. \eqref{pialfa} to be invertible with respect to the
$\dot \xi^\alpha$,
the Hamiltonian $H_{el.}$ of the theory can be obtained
by performing the Legendre
transformation
$$
H_{el.} := \pi_\alpha {\dot \xi^\alpha} - \Lambda =
- {\cal T}^0_0 =
- \menog \ T^0_0 \ ,
$$
and therefore it is numerically equal to the energy density.
Once expressed in terms of the canonical
variables $(\xi^\alpha , \pi_\beta )$
and their spatial derivatives,
$H_{el.}$ generates the Hamiltonian version of the field equations
$$
{\dot f} = \{ H_{el.} , f \}
$$

\pretitolo
{\bf 4. ADM formulation of the Einstein field equations
in elastic media.}
\postitolo

In the present Section we are going to derive
the canonical (Hamiltonian) formulation of General Relativity
coupled to a thermo--elastic medium.
First of all, we shall briefly review the corresponding
ADM formulation for the vacuum case.
\par
Given a ``$3+1$ splitting'' ${\cal M} = \Sigma \times {\bf R}^1$ of
spacetime, describe the initial data on
each initial value surface $\Sigma_t = \Sigma \times \{ t
\}$ by a 3-dimensional, Riemannian metric $q_{ij}$ and the corresponding ADM
momentum $P^{ij}$. The 4-dimensional spacetime metric is therefore
equal to
$$
g_{\mu\nu}=
\left(\matrix{&N_iN^i-N^2  &N_i \cr
&N_j &q_{ij} \cr
}\right) \ ,
$$
where the quantities
$$
\eqalign{
N &:= \frac 1{\sqrt{-\gia 00}}\ ,\cr
N_i &:=\gi 0i\ ,
}
$$
are the {\it lapse function} and the {\it shift vector}, respectively.
The inverse metric reads
$$
g^{\mu\nu}=
\left(\matrix{& -1/N^2 &N^i/N^2 \cr
&N^j/N^2 &q^{ij} -N^iN^j/N^2 \cr
}\right) \ .
\autoeqno{gsu}
$$
The ADM momentum density is defined as
$$
P^{ij} :=\sqrt{\det q} \ (K q^{ij} - K^{ij} )
$$
where $K_{ij}$ is the second fundamental form of $\Sigma$ and $K$ is its
trace. The field equations split into
a ``non--dynamical'' part (the four equations ${\cal G}^{00}=0$ and
${\cal G}^0_k =0$) and a ``dynamical'' part (${\cal G}_{ij}=0$).
The non--dynamical part gives four constraints
for the Hamiltonian system described by $(P^{ij} ,q_{ij} )$
and the quantities $N$ and $N_k$ play the role of Lagrange
multipliers.
The constraints may be written as
$$
\eqalign{
\X &=0\ ,\cr
Y_i &=0\ ,
}
$$
where we have defined the following objects:
$$
\eqalign{
\X &:=\frac 1{16\pi} \left[  R - \frac 1{q}\left(
P^{ij} P_{ij} -\due  P^2 \right) \right]
\ ,\cr
Y_i &:=
\frac 1{8\pi \sqrt{q} } \nabla_k  P^k_i
\ .
}
\autoeqno{chiy}
$$
In the above formulae,
$R$ and $ \nabla$ denote the Ricci scalar and
the covariant derivative with respect to the 3-dimensional metric
$q_{ij}$, respectively, $P$ is the trace of $P^{ij}$ and $q:= \det
q_{ij}$.
\par
As in any constrained Hamiltonian system,
the dynamics is not uniquely defined.
In fact, one has the freedom of fixing
the Lagrange multipliers
$N$ and $N_k$ at each point of $\Sigma$ and at each instant of
``time'' $t = x^0$.
Such a freedom reflects
the gauge invariance of General Relativity
with respect to the group of spacetime diffeomorphisms.
\par
The Hamiltonian equations governing this system can be shortly written
as follows:
$$
-\delta H
=
\frac 1{16\pi}
\int_\Sigma
{\dot{P}}^{ij}
\delta  q_{ij}
-
\dot{q}_{ij} \delta P^{ij}
 \ . \autoeqno{deltah-empty}
$$
This formula is the field-theoretical version of the standard,
finite-dimensional Hamiltonian formula $-dH(p,q) = \dot{p} dq - \dot{q}
dp$. The gauge
properties of the Hamiltonian formulation of General Relativity
are reflected in the fact that
the Hamiltonian vector field $({\dot{P}}^{ij} , \dot{q}_{ij} )$ {\it
is not} uniquely given by the variation of the Hamiltonian.
This happens because {\it not all} the variations $(\delta P^{ij} ,
\delta q_{ij} )$
are allowed in \eqref{deltah-empty} but only those respecting the
constraints. Consequently, $({\dot{P}}^{ij} , \dot{q}_{ij} )$ are not given
uniquely, but only up to vectors ``orthogonal to the constraints''
(in the sense of the
symplectic structure $\int \delta P^{ij} \wedge \delta q_{ij} $).

In formula \eqref{deltah-empty} we have skipped
the usual volume term $(N X + N^i Y_i )$ because
we are going to work ``on shell'', where the constraints
vanish {\it identically}. Consequently,
the quantity $H$ contains only
``surface terms'' (cfr. e.g. Misner et al. 1973).
In the asymptotically flat case
$H$ equals the ADM-mass calculated at space infinity,
while for compact
$\Sigma$ the Hamiltonian vanishes identically and the entire
information about the dynamics may be retrieved from the constraints.
For a discussion of a ``quasi local''
situation, where the mixed ``initial value + boundary
value'' problem in a bounded subset $V \subset \Sigma$ with not trivial
boundary $\partial V$ is considered,
we refer the reader to a recent paper (Kijowski 1997).
It
contains a general formula for the quasi-local Hamiltonian $H$
expressed in terms of a surface integral over $\partial V$.

Coupling gravity to any matter theory consists in supplementing the
above phase space of the gravitational Cauchy data by the Cauchy data for
the matter fields. In the particular case of isentropic
thermo-elasticity this means that the complete phase space will be
described by twenty objects
$(P^{ij} ,q_{ij}, \pi_\alpha , \xi^\alpha)$.
These objects have to fulfil
to the constraints:
$$
\eqalign{
\X &= \frac {N^2}\menog {\cal T}^{00} = \frac N{\sqrt{q}} {\cal T}^{00} \cr
Y_i&= - \frac N{\menog} {\cal T}^0_i =-\frac 1{\sqrt{q}} {\cal T}^0_i \ ,
} \autoeqno{xyt0}
$$
where the matter energy density and the momentum density on the right hand
side are given by \eqref{tenso}.
These quantities
have to be expressed in terms of the canonical
variables. This leads
to an explicit form of
the Gauss -- Codazzi constraints, relating the geometric quantities
$X$ and $Y_i$ with the material quantities (see equations (51) below).
\par
The Hamiltonian formula generating the
dynamics of the system now reads:
$$
-\delta H
=
\frac 1{16\pi}
\int
{\dot{P}}^{ij}
\delta
q_{ij}
-
{\dot{q}}_{ij} \delta
P^{ij}
+\int \dot \pi_\alpha \delta \xi^\alpha -\dot \xi^\alpha \delta
\pi_\alpha \ ,\autoeqno{deltah}
$$
and again it defines uniquely the dotted quantities {\it up to a gauge},
i.~e.~up to the symplectic annihilator of the constraints.

It was recently proved (Kijowski 1997) that the geometric quasilocal
surface integral defining $H$ is {\it universal} in the sense, that the
Hamiltonian it defines is correct for any matter field and for the
empty space, as well.  One {\it should not}, however,
conclude that the dynamics
of the gravitational field coupled to a matter field does not depend
upon the specific properties of the matter. Indeed, for a given matter
field, the Hamiltonian has to be considered as a function defined on the
phase space of Cauchy data. These data must satisfy those {\it specific}
constraints, which are
implied by the {\it specific} properties of the considered material.
There
is no possibility to identify Cauchy data belonging to two
different spaces, corresponding to different theories of matter.
Hence, even if defined by the same boundary integral, the Hamiltonian
corresponding to such two different matter fields generate two
different field dynamics.
In particular, $H$ is
always the ADM-mass in the asymptotically flat case and vanishes
identically in the spatially-compact case.

\pretitolo
{\bf 5. Entropy picture and the
reduction of the theory with respect to constraints.}
\postitolo

Due to the diffeomorphisms invariance of the above described
theory we are allowed
to impose four conditions on the Cauchy data $(P^{ij} ,q_{ij},
\pi_\alpha , \xi^\alpha )$ in order to reduce it with respect to
the constraints. As far as the ``spatial gauge'' is concerned,
it is
somewhat natural to use the comoving frame, defined by the matter itself:
$x^a =\xi^a$. This means that we identify the matter space
${\calZ}$ with our Cauchy space $\Sigma$ and that the velocity vector
has only the time-component:
$$
u^\mu = \frac 1{\sqrt{-g_{00}}} \delta_0^\mu \ . \autoeqno{u-N}
$$
The main idea of the present approach consists in choosing a temporal
gauge in which we identify also the physical time
with the material
time: $x^0=\xi^0$. This 4-dimensional ``comoving gauge'' implies,
therefore, that we have:
$\ixi \alpha\mu = \delta^\alpha_\mu$. Consequently, formula \eqref{u-N}
implies that the gauge condition for the time variable $x^0$ is
equivalent to
$$
T= \beta \ixi 0\mu u^\mu = \frac {\beta}{\sqrt{-g_{00}}}\ .
\autoeqno{T-N}
$$

Physically, the above equation means
that the scale of time is no longer arbitrary but is uniquely fixed by the
temperature of the material.
We stress that,
unlike many
other gauge conditions used in General Relativity to fix the time
variable (e.~g.~maximal surfaces, constant mean curvature etc.),
this
gauge condition does not impose any restriction on the choice of
possible Cauchy surfaces. In particular, we will prove in Section 7
that the physical quantities describing the thermo-mechanical state of
matter do not depend upon the particular choice of the Cauchy surface.

The gauge condition \eqref{T-N} generates an additional volume term
in the Hamiltonian. This is due to the fact that,
in this gauge, we have $\delta \xi^\alpha=0$ and, consequently,
$$
\dot \pi_\alpha \delta \xi^\alpha -\dot \xi^\alpha \delta
\pi_\alpha =-\delta \pi_0\ .
$$
Being a complete variation,
the above quantity may be carried to the
left hand side of \eqref{deltah}.
But formulae \eqref{sigma} and \eqref{star} imply that
$$
\pi_0 =P^0_{\ 0}=\beta S J^0 =\beta S r
\epsilon^{0\nu\rho\sigma}\id 1\nu \id 2\rho \id 3\sigma
=\beta Sr \ .
$$
Hence, the resulting Legendre transformation of \eqref{deltah} gives us
the following generating formula:
$$
-\delta \widetilde{H}
=
\frac 1{16\pi}
\int
{\dot{P}}^{ij}
\delta
q_{ij }
-
{\dot{q}}_{ij} \delta
P^{ij}\ ,\autoeqno{dhtilde}
$$
where the quantity
$$
\widetilde{H}:=H -\beta \int S r \ ,
$$
plays the role of the total Hamiltonian of the system described
by the canonically conjugate variables $(P^{ij} ,q_{ij})$ only.
It contains not only the surface term $H$ but also a non-vanishing
volume term proportional to the total entropy $\int S r$ (molar entropy
$S$ integrated over the material space with respect to its volume
structure $r$).

In the simplest case of a spatially compact spacetimes the quantity $H$
vanishes and the dynamics is governed by the Hamiltonian
$$
U := \beta rS \ ,\autoeqno{brs}
$$
which, due to eq. \eqref{dhtilde},
generates the evolution equations
$$
\eqalign{
{\dot{P}}^{ij}
&=16\pi \frac{\delta U}{\delta q_{ij}}\ ,\cr
{\dot{q}}_{ij}
&=-16\pi \frac{\delta U}{\delta P^{ij}} \ .\cr
} \autoeqno{U}
$$
In the case of a bounded piece of material $V$ with non-vanishing
boundary, the boundary term of the Hamiltonian provides us a tool to
handle the behaviour of the canonical variables on $\partial V$,
according to each specific boundary problem we want to consider.
In fact,
the evolution of the field within $V$ (and the definition of the phase
space of the system), is not complete unless we specify the appropriate
boundary conditions for the fields $(P^{ij}, q_{ij })$ on $\partial V$.
The dynamical equations of the theory are always
given by \eqref{U}, but they become closed only when a specific
boundary value problem -- and, consequently, a specific form of the
boundary term $H$ -- is chosen. For a discussion of the boundary
``phenomena'' we refer to Kijowski (1997).
\par
In order to be able to interpret the entropy as the Hamiltonian of the
composed ``gravity + matter'' system we have first to interpret it as a
thermodynamical generating function in the so called {\it entropy
picture}. This picture is obtained from \eqref{e-term}:
$$
dS(e,G^{ab}) =
\frac 1T \left(de - \frac 12 {Z_{ab}}dG^{ab} \right)\ ,
\autoeqno{desse}
$$
where the constitutive equation $e = e(G^{ab},S)$ has been solved with
respect to the entropy and the latter has been
taken as the generating function, that is:
$$
S=S(e,G^{ab})\ . \autoeqno{S-function}
$$
This function plays the role of the constitutive equation of the material
in the entropy picture and defines {\it via} \eqref{desse} the response
of the material to changes of the control parameters $(e,G^{ab})$ (in
particular, $1 / T$ plays role of the response to changes of $e$).
In the particular case of perfect fluids, the entropy picture is defined
by the well known formula
$$
dS(e, v) =
\frac 1T (de - pdv )\ .
$$
\par
The
function \eqref{S-function} becomes the Hamiltonian of the theory only
once we are able to express its parameters in terms of the canonical
variables data $(P^{ij} , q_{ij })$. For this purpose we treat the four
constraint equations \eqref{xyt0} as implicit definitions of the lapse
and the shift in terms of the four geometric quantities $X$ and
$Y_i$, together with the space metric $q_{kl}$.
Once a specific material (i.~e.~a specific constitutive equation) has
been chosen, all the thermo-elastic control parameters $(e,G^{ab})$
become uniquely defined as functions of the data $(P^{ij} ,q_{ij })$
{\it via} the quantities $X$ and $Y_i$ (together with a possible
{\it direct} dependence upon $q_{kl}$). Indeed, formula \eqref{gsu}
proves that we have
$$
G^{ij} = q^{ij} - V^i V^j \ ,\autoeqno{proves}
$$
where $q^{ij}$ is the inverse space metric, whereas the ``velocity''
$V$ is defined by the lapse and the shift as follows:
$$
V^k := \frac {N^i}N \ .
$$
On the other hand, formula \eqref{T-N} together with the geometric
identity
$$
g_{00} = N_k N^k - N^2 = N^2 ( V_k V^k - 1)
$$
enables us to
express uniquely the temperature in terms of the lapse
and the shift. The problem consists, therefore, in
solving the four constraints (e.~g.~in the form of equations (51)
below)
with respect to the four
unknown quantities $N$, $N_i$. This is only possible for a specific
material, when the constitutive equations (41) are explicitly given.
For each chosen
material these equations uniquely define the lapse and the
shift in terms of the canonical variables. Finally, equation
\eqref{proves} and the constitutive equation enable us to express
$e$ and $G_{ij}$ in terms of the latter.
Inserting their values into
\eqref{S-function} we finally obtain the entropy as a function
($F$, say) of the canonical variables:
$$
F({\X},Y_k,q^{kl}) = S\left(e({\X},Y_k,q^{kl}),
{G}^{ij}({\X},Y_k,q^{kl})\right)\ .\autoeqno{eqo}
$$
This gives us the Hamiltonian $U$
{\it via} formula \eqref{brs}.
\par
The above procedure may be of little use in practice, because for
realistic materials the resulting
constraint equations may be highly non-linear and their analytic
solution practically
impossible to obtain.
As will be explained in the next section,
this difficulty can be circumvented and the Hamiltonian $F({\X},Y_k,q^{kl})$
can be found as solution of a {\it universal} system of differential
equations in 10 variables $({\X},Y_k,q^{kl})$. In this approach, the
constitutive equations of a {\it specific} material enter
only as boundary data on the surface  $\{ Y_k = 0 \} $.
\par
We stress that in the present picture the canonical variables $(P^{ij}
,q_{ij })$ {\it are not constrained}. They carry the information about 6
independent degrees of freedom of the physical system under
consideration: 2 for gravity and 4 for thermo-elasticity.
Equations \eqref{xyt0} are no longer constraints: they allow us to
reconstruct the lapse and the shift and, consequently,
all the remaining physical quantities characterizing both
the gravitational and the thermo-elastic fields, in terms of the
canonical variables.

With respect to the vacuum case, the above theory consists in
replacing the vanishing Hamiltonian $U \equiv 0$ on the constraint
subspace $X=0$ and $Y_i = 0$ by a non-trivial Hamiltonian $U =
U(X,Y_i,q^{ij})$ and in relaxing completely the constraints. The dynamical
equations generated this way for the quantities $(P^{ij} ,q_{ij })$
carry not only the dynamics of the gravitational field,
but also that of the matter
coupled to gravity.

We are going to prove in the sequel that the theory of
empty space may be obtained as a limiting case of theories with
non-trivial matter, when the density of matter tends to zero. For this
purpose let us consider a family of state equations
$$
e_c ( G^{ab} , S) = c e( G^{ab} , S) \ , \autoeqno{rescaled}
$$
where $c$ is a positive constant and $e=e( G^{ab} , S) $ corresponds to a
reference material. The material described by the new state equation
\eqref{rescaled}
differs from the reference material in the following way: the total
mass of a piece of the {\it new material} is $c$ times the mass of
the same piece of the {\it reference material}
(by {\it the same} piece we mean that it is in the same state of strain
$G^{ab}$ and contains the same amount of entropy). We will
prove at the end of the next Section that the rescaled state equation
\eqref{rescaled} leads to the
following Hamiltonian, when the material is coupled to gravity:
$$
U_c (X,Y_i,q^{ij}) := U(\frac X{c} ,\frac {Y_i}c ,q^{ij}) \ .
\autoeqno{Uc}
$$
The limit $c \rightarrow 0$ corresponds to a very light matter. In this
regime the values of $U_c$ become very big outside of the subspace
$\{X=0 ; Y_i = 0\}$ and remain bounded only on the constraints. This
way the constraints arising in the vacuum case may be considered as a
limiting case of a {\it very deep} ``potential well'', corresponding to
a {\it very light} matter.

\pretitolo
{\bf 6. Structure of the Hamiltonian.}
\postitolo

Different materials are characterized by different Hamiltonian
$U=\beta r S$.
However, not all the functions of the
ten parameters $({\X},Y_k,q^{kl})$ may be obtained
from an arbitrarily chosen constitutive function \eqref{S-function} of
seven parameters through the Legendre transformation described above.
Indeed, the function $U$
has the following, universal properties:
\capo
{{\bf Theorem 1}}
\capo
{\it 1) The function $U$ fulfils the following system of three first-order
partial differential equations:
$$
2 \diff U{Y_l} \diff U{q^{kl}}
=
\left(\diff U\X \right)^2
Y_k\ .\autoeqno{fine}
$$
\capo
2) For vanishing $Y_i$, the shift vector vanishes and
the function $U$ satisfies the following boundary condition:
$$
U(X,0,0,0, q^{kl}) = \beta rS\left(\frac {\X}{r\sqrt{\det q^{ij}}},
q^{kl}\right) \ ,
\autoeqno{initial}
$$
where $S$ is the constitutive function
\eqref{S-function}
of the material.
\capo
3) The above boundary value problem for equations
\eqref{fine}
is well
posed
and may be solved by the characteristics method.
\capo
4) The lapse function $N$, and the shift vector $N^k$
are uniquely given by the derivatives of $U$
according to the following formulae:
$$
\eqalign{
N &=\frac 1{\sqrt{q}} \diff U\X \ ,\cr
N^l &=\frac 1{\sqrt{q}}  \diff U{Y_l}\ .
}\autoeqno{nnk}
$$
}
\capo
{\it Proof:}
\capo
Due to equality \eqref{eqo}
we have
$$
\eqalign{
\frac 1{\beta r}
\diff U{\X} &= \diff Se \diff e{\X} + \diff S{{G}^{ij}}
\diff {{G}^{ij}}{\X}\ ,
\cr
\frac 1{\beta r}
\diff U{Y_k} &= \diff Se \diff e{Y_k} + \diff S{{G}^{ij}}
\diff {{G}^{ij}}{Y_k}\ ,
\cr
\frac 1{\beta r}
\diff U{q^{kl}} &= \diff Se \diff e{q^{kl}}
+ \diff S{{G}^{ij}} \diff {{G}^{ij}}{q^{kl}}
\ .
}\autoeqno{deffea}
$$
Now consider the constraints \eqref{xyt0}.
Using formula \eqref{tenso} and recalling that $\ixi a\mu =\id a\mu$
in our gauge, these read:
$$
\eqalign{
X&=\frac 1v \left(\frac e{1-V^2} + Z_{kl} V^k V^l \right)\ ,\cr
Y_k&=-\frac 1v
\left(\frac e{1-V^2}V_k + Z_{kl}V^l \right)\ ,
}\autoeqno{muk1}
$$
where $V^2=q_{kl}V^k V^l$.
These equations may be rewritten as
$$
\eqalign{
e &=v\left(X+Y_kV^k\right)\ ,\cr
Z_{kl}V^l &=-\left(
vY_k + \frac e{1-V^2}V_k\right)\ .
}
\autoeqno{muk}
$$
Due to formula \eqref{desse}, one has
$$
\diff S{G^{kl}}
=-\due {\diff Se} Z_{kl}\ .
$$
Using it and
\eqref{proves},
we obtain
$$
\eqalign{
\diff S{{G}^{ij}}
\diff {{G}^{ij}}{\X} &=
Z_{ij}V^j \diff {V^i}{\X} \diff Se=
-\left(vY_i+\frac e{1-V^2}V_i\right)\diff {V^i}{\X} \diff Se
\ ,
\cr
\diff S{{G}^{ij}}
\diff {{G}^{ij}}{Y_k} &=
Z_{ij}V^j \diff {V^i}{Y_k}\diff Se
=
-\left(vY_i+\frac e{1-V^2}V_i\right)\diff {V^i}{Y_k}\diff Se
\ ,\cr
\diff S{{G}^{ij}}
\diff {{G}^{ij}}{q^{kl}} &=\left(
\diff S{{G}^{kl}} + Z_{ij}V^j \diff {V^i}{q^{kl}}\right)\diff Se
=\left[\diff S{{G}^{kl}}
-\left(vY_i+\frac e{1-V^2}V_i\right)\diff {V^i}{q^{kl}}\right]
\diff Se
\ .\cr
}
$$
Now we calculate
the derivatives of the function $e=e(\X ,Y_k ,q^{kl})$
using the first equation of \eqref{muk}
and the definition of $v$ (see \eqref{numero}):
$$
v=\frac 1\rho =\frac 1{r\sqrt{\det (q^{kl}-V^k V^l)}}=
\frac 1{r\sqrt{\det q^{kl}}
\sqrt{1-V^2}
}\ .\autoeqno{v1r}
$$
This way we obtain
$$
\eqalign{
\diff e{\X} &=
v+\left(vY_i+\frac e{1-V^2}V_i\right)\diff {V^i}{\X}
\ ,\cr
\diff e{Y_k} &=
v V^k +
\left(vY_i+\frac e{1-V^2}V_i\right)\diff {V^i}{Y_k}
\ , \cr
\diff e{q^{kl}} &=
-\frac e2 \left(q_{kl}+\frac {V_k V_l}{1-V^2}\right)
+
\left(\frac {eV_j}{1-V^2}+
vY_j\right)\diff {V^j}{q^{kl}}\ .
}
$$
Inserting the above results in eqs. \eqref{deffea}
we obtain:
$$
\eqalign{
\frac 1{\beta r}
\diff U{\X} &= v \diff Se \ ,
\cr
\frac 1{\beta r}
\diff U{Y_k} &= v \diff Se {V^k}
\ ,
\cr
\frac 1{\beta r}
\diff U{q^{kl}} &=
\diff S{G^{kl}}
-\due
\diff Se e
\left(q_{kl}+\frac {V_kV_l}{1-V^2}\right)\ .
}
\autoeqno{deffea1}
$$
Contracting the last equation with $V^k$ and using
the other two together with the vector constraint \eqref{muk},
we finally obtain \eqref{fine}.
\par
To prove the validity of the initial condition \eqref{initial} we
observe that
for $Y_k=0$ the scalar constraint in \eqref{muk} reduces to $e=v\X$,
whereas the vector constraint gives us
$$
\left(\frac e{1-V^2}q_{kl} + 2Z_{kl} \right) V^l =0\ .
$$
For generic equations of state the tensor
on the left is non--singular
(it becomes singular only
if one of the principal stresses - eigenvalues of the stress tensor -
equals the large negative value $-e\rho/2(1-V^2)$).
Hence, in a generic situation $V^l$ must vanish on initial data.
Using \eqref{v1r} with $V=0$ to express $v$ in $e=v\X$,
we finally obtain \eqref{initial}.
\par
To prove the integrability of the system
composed by the three equations \eqref{fine},
we denote
$$
P_X:=\diff {U}\X \ ,P^i:=\diff U{Y_i}\ ,\Pi_{ij}:=\diff U{q^{ij}}
$$
and rewrite eqs. \eqref{fine} as three Hamilton--Jacobi equations
$$
{\bf H}_k\left(\X, Y_i , q^{ij} , P_X , P^i , \Pi_{ij} \right)=0 \ ,
\autoeqno{bfg}
$$
where the functions
$$
{\bf H}_k:=2\Pi_{kl}P^l -P_X^2 Y_k\ ,
$$
may be viewed as Hamiltonian defined
on a 20--dimensional phase space ${\cal P}$ parameterized by the
coordinates $Q_\Sigma ,P^\Sigma $, where
$\Sigma =1,..,10$ and
$$
Q_\Sigma = (\X ,Y_i ,q^{ij})\ .
$$
On this phase space define the ordinary Poisson bracket as
$$
\{ {\bf F},{\bf G}\}
=\sum_{\Sigma=1}^{10} \left(\diff {{\bf F}}{Q_\Sigma}\diff {{\bf
G}}{P^\Sigma} -
\diff {{\bf G}}{Q_\Sigma}\diff {{\bf F}}{P^\Sigma} \right)\ .
$$
Now, it is easy to check that
$$
\{ {\bf H}_k ,{\bf H}_l \} = 0\ .
$$
This means
that the three dynamical systems
are in involution.
\par
To prove point 3 of the theorem, we will propagate the initial value
\eqref{initial} of the function $U$ over the characteristic lines of the
three Hamiltonian.  For this purpose we first
calculate the values of the momenta $P_X$ and $\Pi_{ij}$ on the initial
surface $\{ Y_k = 0 \}$ from the derivatives of the entropy
\eqref{initial} with respect to $X$ and $q^{ij}$. Then we solve
the equations \eqref{bfg} algebraically, with respect to the
remaining three
momenta $P^i$. This way we obtain, at each point of the initial surface,
the complete set of initial data for the trajectories of the three
Hamiltonian. The collection of all these data defines a 7-dimensional
surface in the phase space ${\cal P}$. Because the Hamiltonian are in
convolution, the trajectories starting from each point of the surface
span a 3-dimensional characteristic subspace. The method of
characteristics tells us that the function $U$ must
be constant on these subspaces (see e.g.
Courant \& Hilbert 1989).
The collection of all the characteristic subspaces
forms a 10-dimensional Lagrangian submanifold ${\cal D}$ of ${\cal P}$
and the function $U$ is defined on ${\cal D}$. The solution of the
problem is then obtained by projecting this function from ${\cal D}$ down to the
``configuration space'' of the parameters $Q_\Sigma$.
\par
Finally, to prove the last part of the Theorem, we observe that,
due to \eqref{T-N} and \eqref{v1r}, we have:
$$
v\diff Se =\frac vT =
\frac {v\sqrt{-g_{00}}}{\beta}
=\frac {N \sqrt{q}}{\beta r}
$$
and thus the first two equations of \eqref{deffea1} reduce to \eqref{nnk}.
\c
\capo
{\it Remark}
\capo
The projection of the Lagrangian submanifold ${\cal D}$
to the configuration space of the parameters $Q_\Sigma$
may become singular on caustic surfaces.
It was proved in KSG that,
at least in the case of fluids, convexity of initial data
\eqref{initial}, implied by the physical properties of the entropy
function, excludes the existence of
singularities and implies that the function $U$ may
be always constructed {\it globally}.
In the case of a generic material this problem
needs further investigations.
\par
Finally, the following corollary of the previous Theorem shows how to
reconstruct the vacuum gravity theory as a limiting case of the present
theory,
when matter becomes very light:
\capo
{\bf Corollary}
\capo
{\it If $U$ is a solution of equations}
\eqref{fine}
{\it , the function $U_c$ defined by formula}
\eqref{Uc}
{\it also satisfies the
same equations and, therefore, may be taken as a possible Hamiltonian of the
theory. It describes the dynamics of the material corresponding to the
rescaled state equation}
\eqref{rescaled}.
\capo
{\it Proof}:
\capo
The first statement may be easily checked by inspection.
Moreover, let us observe that the rescaling \eqref{rescaled} of the energy
is  equivalent to the following rescaling in the entropy picture:
$$
S_c ( e, G^{ab} ) = S( \frac ec , G^{ab} ) \ .
$$
To prove that this relation is indeed satisfied by the material
corresponding to the new Hamiltonian,  consider the initial data
\eqref{initial} for $U_c$:
$$
\eqalign{
S_c ( e, q^{kl} ) &= \frac 1{\beta r}U_c(X,0,0,0, q^{kl}) = \frac
1{\beta r}
U(\frac Xc ,0,0,0, q^{kl}) = \cr
&=
S\left(\frac {\X}{cr\sqrt{\det q^{ij}}},
q^{kl}\right) = S(\frac ec , q^{kl} ) \ ,
}
$$
which ends the proof.
\c

\pretitolo
{\bf 7. Gauge invariance of the entropy.}
\postitolo

Physically, the Hamilton--Jacobi conditions \eqref{fine}, imposed on
the possible Hamiltonian $U$ are equivalent to the
invariance of the entropy with respect to spacetime diffeomorphisms. In
fact, the function $U$ ``has to derive from its arguments'' $(X,Y_k,
q^{kl})$ the amount of the purely material quantity $S$. Performing a
spacetime diffeomorphisms we may change completely the data $(X,Y_k,
q^{kl})$. However, the value of $U$ assigned to the new data must
remain the same as before, since the amount of entropy contained in
the material does not depend upon the parameterization
of the initial data.

The invariance of the entropy with respect to purely 3-dimensional
diffeomorphisms of ${\calZ}$ (space diffeomorphisms) is automatically
satisfied due to the fact that both the data $(X,Y_k, q^{kl})$ and the
material structure defining the state equation are geometric objects
defined on the matter space, whereas $S$ is a scalar. Hence, only
diffeomorphisms changing the time variable may be
dangerous from this point of view. These ``generalized boost
transformations'' correspond to non trivial changes of the Cauchy
surface in the spacetime.
We are going to prove that they also do not change the value of $U$.

In fact, consider a transformation which reduces to the identity on
${\calZ}$ and consists in the translation of the material time
$$
x^0 \longrightarrow x^0 + \varphi (x^k ) \ .
$$
This transformation may be treated as generated by the Hamiltonian
$$
U_\varphi (X,Y_k,q^{kl}):= \varphi (x^k ) \ U(X,Y^k,q^{kl}) \ .
$$
Hence, we are going to prove that, for any function $\varphi$ defined
on ${\calZ}$, the dynamics generated by $U_\varphi$ preserves the value
of $U_\varphi$ {\it at each point separately}, the global invariance
being obvious because the integral of $U_\varphi$, i.~e.~the Hamiltonian,
is always conserved by its own dynamics.
In particular, for $\varphi
\equiv 1$ we obtain the local entropy conservation with respect to
the dynamics discussed previously.

\capo
{{\bf Theorem 2 }}
\capo
{\it The equations
\eqref{fine}
imply the local conservation of entropy with
respect to the dynamics generated by $U_\varphi$.}
\capo
{\it Proof:}
\capo
To simplify the proof it is convenient to introduce
the vector density associated to the vector $Y_i$:
$$
y_i := \sqrt{q} \ Y_i =
\frac 1{8\pi } \nabla_k  P^k_i\ , \autoeqno{maley}
$$
and to express the Hamiltonian $U_\varphi$ in terms of this variable
and the {\it covariant} (instead of the contravariant) metric. We
denote:
$$
W_\varphi (X,y_k,q_{kl}):=\varphi (x^l )\ U\left(
X,\frac{y_k}{\sqrt{q}} , q^{kl}\right)\ .
$$
First of all, we are going to derive the evolution equations \eqref{U} with
the Hamiltonian $U$ replaced by $W_\varphi$. For this purpose we
calculate the total variation $\delta {W_\varphi}$:
$$
\delta {W_\varphi} = \diff {W_\varphi}\X \delta \X +
\diff {W_\varphi}{y_k} \delta y_k +
\diff {W_\varphi}{q_{kl}}\delta q_{kl}\ .\autoeqno{du}
$$
Using the definition \eqref{chiy}, we obtain
$$
16 \pi \delta \X =
\delta R +\frac 1q \left(P^{ij}P_{ij} -\frac
{P^2}{2}\right)q^{kl}\delta q_{kl}
-\frac 1q \left[
(2P_{kl} -Pq_{kl})\delta P^{kl} + (2P^k_mP^{ml}-PP^{kl})\delta q_{kl}
\right].
$$
The variation of the Ricci scalar may be written as follows:
$$
\delta R = -R^{kl}\delta q_{kl} + (\id ls q^{mn}-\id ns q^{ml})
\Gamma^k_{lk}\delta\Gamma^s_{mn} +\dib l \left[
(\id ls q^{mn}-\id ns q^{ml})\delta\Gamma^s_{mn}\right]\ ,
$$
therefore we obtain:
$$
\eqalign{
16 \pi \delta \X
&=
\left[
-R^{kl}
+\frac 1q \left(P^{ij}P_{ij} -\frac {P^2}{2}\right)q^{kl}
-\frac 1q (2P^k_mP^{ml}-PP^{kl})\right]
\delta q_{kl} +
\cr
&-\frac 1q
(2P_{kl} -Pq_{kl})\delta P^{kl}
+
(\id ls q^{mn}-\id ns q^{ml})
\Gamma^k_{lk}\delta\Gamma^s_{mn} +\dib l \left[
(\id ls q^{mn}-\id ns q^{ml})\delta\Gamma^s_{mn}\right]\ .
}
\autoeqno{dx}
$$
Due to \eqref{maley} we have:
$$
8\pi \delta y_k =
\dib l (P^{lm}\delta q_{km} +
q_{km}\delta P^{lm})-\due (P^{lm}\delta q_{lm,k}
+ q_{lm,k} \delta P^{lm} )\ .\autoeqno{dy}
$$
Inserting formulae \eqref{dx} and \eqref{dy}
in \eqref{du}, collecting the terms containing independent
variations and eliminating all the boundary terms (total divergences)
we finally obtain the following evolution equations generated by
$W_\varphi$:
$$
\eqalign{
\dot q_{kl}
= &
\frac 2q \diff {W_\varphi}\X
\left(P_{kl}-\due P q_{kl}\right)
+q_{km} \nadib l \diff {W_\varphi}{y_m}
+q_{lm} \nadib k \diff {W_\varphi}{y_m}
\ ,
\cr
\dot P^{kl}
= &
\diff {W_\varphi}\X \left[
-R^{kl}
+\frac 1q \left(P^{ij}P_{ij} -\frac {P^2}{2}\right)q^{kl}
-\frac 1q (2P^k_mP^{ml}-PP^{kl})\right]
+ \cr
&
+
\nadiba k\nadiba l \diff {W_\varphi}\X - q^{kl}
\nadib m\nadiba m \diff {W_\varphi}\X
- P^{km}\nadib m \diff {W_\varphi}{y_l}
- P^{lm}\nadib m \diff {W_\varphi}{y_k}
+
\cr
&+
\nadib m \left(
P^{kl}\diff {W_\varphi}{y_m}\right)
+ 16 \pi \diff {W_\varphi}{q_{kl}}\ ,
}
\autoeqno{evo}
$$
(the above formulae may be rewritten in a somewhat more familiar
form if we replace the derivatives of $W_\varphi$
with respect to $X$ and $y_l$
by the lapse and the shift,
using eqs. \eqref{nnk}).
To calculate $\dot {W_\varphi}$,
we may simply rewrite formulae \eqref{dx} and \eqref{dy}, replacing
the variations of $q_{kl}$ and $P^{kl}$ by their time derivatives.
This way we obtain:
$$
\eqalign{
16 \pi \dot \X
&
=
\left[
-R^{kl}
+\frac 1q \left(P^{ij}P_{ij} -\frac {P^2}{2}\right)q^{kl}
-\frac 1q (2P^k_mP^{ml}-PP^{kl})\right]
\dot q_{kl} +
\cr
&-\frac 1q
(2P_{kl} -Pq_{kl})\dot P^{kl}
+
\nadib l \left(
\nadib m \dot q^{ml} -\nadiba l \dot q^m_m \right)\ ,\cr
8\pi \dot y_k
&
=
\nadib l (P^{lm}\dot q_{km} +
q_{km}\dot P^{lm})-\due P^{lm}\nadib k q_{lm}
\ .}
\autoeqno{dotxy}
$$
Finally, inserting \eqref{evo} into \eqref{dotxy} we obtain:
$$
\eqalign{
\dot \X
&=
\diff {W_\varphi}{y_l}
\nadib l \X +\frac 2{q} y^l\nadib l \diff {W_\varphi}\X +
\frac 1{q} \diff {W_\varphi}\X \nadib l y^l
-\frac 2{q}
\left(
P_{kl} -\due P q_{kl}
\right)
\diff {W_\varphi}{q_{kl}}
\ ,
\cr
\dot y_k
&=
-\diff {W_\varphi}\X
\nadib k X +
\nadib l \left(y_k
\diff {W_\varphi}{y_l}
\right)
+
y_l \nadib k
\diff {W_\varphi}{y_l}
+2\nadib l
\left( q_{km} \diff {W_\varphi}{q_{lm}} \right)\ .
}
$$
Plugging the above results into formula:
$$
\dot {W_\varphi} =\diff {W_\varphi}\X \dot X +
\diff {W_\varphi}{y_k}\dot y_k +\diff {W_\varphi}{q_{kl}}\dot q_{kl}\ ,
$$
one may readily check  the following result:
$$
\dot {W_\varphi} =\nadib l \left[
y_k \diff {W_\varphi}{y_k} \diff {W_\varphi}{y_l}
+ \frac 1q \left(\diff {W_\varphi}\X \right)^2 q^{lk}y_k +2
\diff {W_\varphi}{y_k}
q_{km}\diff {W_\varphi}{q_{ml}}
\right]\ .\autoeqno{dotu}
$$
Now, we
come back to our variables $Y_i$ and to the function $U$.
For this purpose we observe that
$$
\eqalign{
\diff{W_\varphi}X &=\varphi \diff UX\ ,\cr
\diff{W_\varphi}{y_k}&=\frac 1{\sqrt{q}} \varphi \diff{U}{Y_i}\ ,\cr
\diff{W_\varphi}{q_{kl}}
&=\varphi\left(-q^{ik}q^{jl}\diff U{q^{ij}}
+\due Y_i \diff U{Y_i}\right)\ .
}
$$
Plugging these into formula \eqref{dotu} we finally obtain
$$
\varphi \dot U =
 \dib l Q^l \ ,
$$
where we have defined
$$
Q^l :=
q^{lm}
\frac {\varphi^2}{\sqrt{q}}
\left[
\left(\diff U\X \right)^2 Y_m
-2
\diff U{Y_k}
\diff U{q^{km}}
\right]
\ .\autoeqno{defsigma}
$$
This vector density vanishes
due to the Hamilton--Jacobi equations \eqref{fine}.
This ends the
proof. \c
\par
Physically, the vector density $Q^l$
is equal to the
entropy current (i.~e.~to the heat flow). We conclude that the
Hamilton--Jacobi conditions imposed on $U$ are a consequence
of the fact
that we are considering only isentropic
phenomena, for which the heat flow vanishes identically.

\pretitolo
{\bf 8. Isotropic elastic media.}
\postitolo

In the case of an isotropic material,
the function of state depends, besides of $e$, only
on the three control parameters $(v,h,q)$ (cfr. section 2, example 2).
Such parameters contain also the material metric which is {\it
a priori} given for each material. Hence, the solution depends also
upon $\gamma_{ij}$, but the
Hamiltonian must be invariant with respect to local (i.~e.~at each
point independently) isometries of $\gamma$. This implies that $S$ may
depend, besides of the scalar $X$, only upon the invariants of
the matrix $\chi^i_{\ j} := q^{ik} \gamma_{kj}$ and upon invariants built
of the vector $Y_i$.  We choose the following set for the
invariants of
$\chi$:
$$
\eqalign{
\Z &:= \det \chi  \ ,\cr
\H &:=  {\rm Tr} \chi \ , \cr
\L &:= {\rm Tr} \chi^2 -({\rm Tr} \chi )^2
}\autoeqno{def-inva}
$$
For the remaining invariants we choose the lengths of $Y_i$
calculated with respect to three different metric tensors:
$$
\eqalign{
\r & := \gamma^{ij} Y_i Y_j  \ , \cr
\s & := q^{ij} Y_i Y_j \ , \cr
\t & := q^{ij} q^{kl} \gamma_{jl} Y_i Y_j \ .
} \autoeqno{def-small-s}
$$
It is, therefore, obvious that the function
$S$ will depend upon $X$, $Y_i$ and $q^{kl}$
{\it via} seven invariants only
$$
S = F(\X , \Z , \H, \L , \r, \s, \t) \ .\autoeqno{F}
$$
The
equations \eqref{fine}
may be rewritten in terms of the above invariants
in the following way:
$$
\eqalign{
\m\Z \m{\r}
+2\m\L\m\s +
\m\H\m\t &=0 \ ,
\cr
\m\t (\r\m\r +\s\m\s +\t\m\t )
&=
\m\t \left( \Z\m\Z +
\H\m\H + \L\m\L \right)
+2 \m\L\m\r +\m\s \m\H
\cr
\due
\left(\m\X\right)^2 -2\m\s (\r\m\r +\s\m\s +\t\m\t )
&
=
2
\left(
\m\H -2\H\m\L
\right)
\m\r
+
\left(
\L \m\H + 4\Z \m\L \right)
\m\t
+
\cr
+
2
\left(\L \m\L +\Z\m\Z \right)
\m\s
&
+
\frac 13
\m\t
\left\{
2\s\m\r
+2\t\m\s
+\left[
2\H \t +2\Z\r
+\L\s
\right]
\m\t
\right\}
\ ,
}\autoeqno{fine2}
$$
where by $p$ with a subscript we denote the derivative of $F$
with respect to the corresponding variable,
e.g. $\m\X =(\partial F/\partial \X )$.
\par
For vanishing $Y_i$, i.~e.~for $\r = \s= \t= 0$, the shift
vector vanishes and, therefore, the tensor $\chi$ coincides with the
strain $h$. Consequently, its invariants $(\Z, \H, \L )$
may be calculated in terms of the invariants $(v,h,q)$ of the strain.
Moreover, the
Hamiltonian constraint still gives $X = e/v $. This implies that the
function $F$ satisfies the following boundary condition:
$$
F(X, \Z , \H, \L ,0,0,0) = S(X/\sqrt{\Z} , 1/\sqrt{\Z}, \H, \L +\H^2 ) \ ,
\autoeqno{initial1}
$$
where $S$ is the constitutive function of the material.
\par
Finally, we are going to discuss two simple examples,
which correspond to particularly simple choice of the equation of state:
\capo
{\it 1. Perfect fluids}
\capo
For a perfect fluid the equation of state depends,
besides of $e$, only on
the specific volume.
Therefore $\m\H =\m\L =0$ identically.
Then the first equation in \eqref{fine2}
gives $\m\r =0$ while the second one then implies
$\m\t (\s\m\s +\t\m\t )
=\m\t \Z\m\Z$.
Choosing the solution $\m\t =0$ we end up
with a function $F=F(\X ,\Z ,s)$
which has to satisfy the partial differential equation
$$
\left(\diff FX \right)^2 -4\s \left(\diff F\s \right)^2
=4\Z \diff F\Z \diff F\s \ .
$$
The above equation
simplifies considerably if put
$$
\Y =\frac s\Z \ ,
$$
in fact in such a case we have (KSG):
$$
\left(\diff FX \right)^2 -4\diff F\Y \diff F\Z = 0 \ .\autoeqno{kung}
$$
\capo
{\it 2. An elastic material resembling a perfect fluid}
\capo
Consider an elastic material for which the state
equation, besides of $e$, depends only on the trace of the
strain tensor.
Physically, this material has always the same response to all the
strains that change its volume without changing
its shape.
Therefore, it may be considered as a ``counterpart'' of the
perfect fluid (the response of a perfect fluid to strains that
do change its shape without changing its volume vanishes).
If $F$ depends on $\H$ only,
we have $\m\Z =\m\L=0$ and
the first equation in \eqref{fine2}
gives $\m\t =0$.
The second one then implies
$\m\s=0$.
Therefore we end up
with a function $F=F(\X ,\H ,\r)$
which has to satisfy the partial differential equation
$$
\left(\diff FX \right)^2 -4\diff F\r \diff F\H = 0 \ .
$$
The above equation is formally identical with
the perfect--fluid equation \eqref{kung}.
\par
For a more detailed discussion of the isotropic case,
see Iacoviello et al. (1996).

\vfill\eject
\pretitolo
\centerline{\bf References}
\postitolo

\item{\ }
Anile, A.M., Choquet--Bruhat, Y., (ed.) (1989)
{\it Relativistic fluid dynamics},
Lecture Notes in Math. {\bf 1385}, Springer Verlag, New York.\capo

\item{\ }
Arnowitt, R., Deser, S., Misner, C.W.  (1962)
The dynamics of general relativity.
In {\it Gravitation:  An introduction to current research,} L. Witten ed.,
Wiley, New York, p. 227--265.
\capo
\item{\ }
Bressan, A., (1978)
{\it Relativistic theories of materials,}
Springer-Verlag, Berlin-New York.
\capo
\item{\ }
Carter, B., Quintana, H. (1972)
Foundations of general relativistic high pressure
elasticity theory.
Proc. R. Soc. Lond. A{\bf 331}, p. 57--83.\capo
\item{\ }
Carter, B. (1973)
Elastic perturbation
theory in general relativity
and a variation principle for a rotating solid star.
Comm. Math. Phys. {\bf 30} p. 261--286.
\capo
\item{\ }
Cattaneo, C. (1973)
Elasticit\'e Relativiste.
Symp. Math. {\bf 12}, Acad. Press NY, p. 337--352.\capo
\item{\ }
Comer, G.L., Langlois, D. (1993)
Hamiltonian formulation for multi-constituent relativistic perfect fluids.
Class. Quantum Grav. {\bf 10}, p.2317--2327.
\item{\ }
Comer, G.L., Langlois, D. (1994)
Hamiltonian formulation for relativistic superfluids.
Class. Quantum Grav. {\bf 11}, p. 709--721.
\item{\ }
Courant, R., Hilbert, D., (1989)
{\it Methods of mathematical
physics}.
Wiley \& Sons ed., New York.
\item{\ }
De Witt, B. (1962)
The quantization of geometry.
In {\it Gravitation: an introduction to current research},
L. Witten ed., Wiley, New York, p.266--381.\capo
\item{\ }
Glass, E.N., Winicour, J. (1972)
General Relativistic elastic systems.
J. Math. Phys. {\bf 13}, p.1934--1940.\capo
\item{\ }
Haensel, P., (1995)
Solid interiors of neutron stars and gravitational radiation,
In {\it Astrophysical sources of gravitational radiation},
J.A. Marck and J.P. Lasota ed. (Les Houches 1995).
\item{\ }
Hernandez, W.C. (1970)
Elasticity in general relativity.
Phys. Rev. D {\bf 1}, p. 1013--1017.\capo
\item{\ }
Hajicek, P., Kijowski, J. (1998)
Lagrangian
and Hamiltonian formalism for discontinuous fluid and gravitational field
Phys. Rev. D {\bf 57}, p.914-935.\capo
\item{\ }
Holm, D. (1989)
Hamilton techniques for relativistic fluid dynamics
and stability theory.
In {\it Relativistic fluid dynamics},
A.M. Anile and Y. Choquet Bruhat ed.,
Lecture Notes in Math. {\bf 1385}, Springer Verlag, New York,
p.65--151.\capo
\item{\ }
Iacoviello, D., Kijowski, J., Magli, G. (1997)
The dynamical structure of the ADM equations
for general relativistic, isotropic elastic media.
In {\it General Relativity and Gravitational Physics},
M. Bassan, F. Fucito, I. Modena, V. Ferrari
\& M. Francaviglia ed.,
W.S. Publ. Co.,
Singapore, p.341-346.
\capo
\item{\ }
Jezierski, J., Kijowski, J. (1991)
Thermo-hydrodynamics as a field theory.
In {\it Hamiltonian Thermodynamics}, editors S. Sieniutycz and P.
Salamon, Taylor and Francis Publishing Company.\capo
\item{\ }
Kijowski, J., (1997)
A simple derivation of canonical structure and quasi--local
Hamiltonian in General Relativity.
Gen. Rel. Grav.
{\bf 29}, p.307--343.
\item{\ }
Kijowski, J., Magli, G. (1992)
Relativistic elastomechanics as a lagrangian field theory.
Geom. and Phys. {\bf 9}, p. 207--223.
\capo
\item{\ }
Kijowski, J., Magli, G. (1997)
Unconstrained variational principle and canonical
structure for relativistic elasticity theory.
Rep. Math. Phys {\bf 39}, p. 99--112.
\capo
\item{\ }
Kijowski, J., Pawlik, B., Tulczyjew W.M. (1979)
A variational formulation of non-gra\-vi\-ta\-ting and gravitating
hydrodynamics, Bull. Acad. Polon. Sci. {\bf 27}  p. 163--170.
\capo
\item{\ }
Kijowski, J., Sm\'olski, A., G\'ornicka A. (1990)
Hamiltonian theory of self--gravitating perfect fluids
and a method of effective deparameterization
of Einstein theory of gravitation.
Phys. Rev. D {\bf 41}, p. 1875--1884.
\capo
\item{\ }
Kijowski, J., Tulczyjew W.M. (1979)
{\it A symplectic framework for field theories}.
Lecture Notes in Physics {\bf 107}, Springer, Berlin.
\item{\ }
Kijowski, J., Tulczyjew, W.~M. (1982) Relativistic hydrodynamics
of isentropic flows, Mem. Acad. Sci.
Torino, serie V, {\bf 6} p. 3--17.
\item{\ }
Kuchar, K.V. (1993)
Canonical quantum gravity.
In ``General relativity and gravitation - Cordoba 1992'' p.119--150
(Inst. Phys. Publ., Bristol, U.K.).
\item{\ }
K\"unzle, H.P., Nester, J.M. (1984)
Hamiltonian formulation of gravitating perfect fluids and the
Newtonian limit.
J. Math. Phys. {\bf 25}, 4, p. 1009--1018.
\capo
\item{\ }
Maugin, G. A., (1971)
Magnetized deformable media in general relativity.
Ann. Inst. H. Poincar\'e' {\bf A15} p.275--302.
\capo
\item{\ }
Maugin, G.A. (1977)
Infinitesimal discontinuities
in initially stressed relativistic elastic solids.
Comm. Math. Phys. {\bf 53}, p.233--256.
\capo
\item{\ }
Maugin, G.A., (1978a)
On the covariant equations of the relativistic electrodynamics of
continua III. Elastic solids.
J. Math. Phys. {\bf 19}, p. 1212--1219.
\capo
\item{\ }
Maugin, G. A., (1978b)
Exact relativistic
theory of wave propagation in prestressed nonlinear elastic solids.
Ann. Inst. H. Poincar\'e' {\bf A28} p.155--185.
\capo
\item{\ }
Maugin, G. A., (1993)
{\it Material inhomogeneities in elasticity,}
Chapman \& Hall, London.
\capo
\item{\ }
Shapiro, S.L., Teukolsky, S.A. (1983)
{\it Black Holes, White Dwarfs and Neutron Stars}.
Wiley, New York.\capo
\item{\ }
Souriau, J.M. (1964)
{\it G\'eom\'etrie et Relativit\'e}.
Hermann, Paris.\capo
\item{\ }
Truesdell, C., Toupin, R., (1960)
{\it The classical field theories,}
In ``Handbuch der Physik,'', Bd. III/1, pp. 226--793;
Springer, Berlin.\capo
\end